# Optimal configuration of cooperative stationary and mobile energy storage considering ambient temperature: A case for Winter Olympic Game

*He Meng, Hongjie Jia, Tao Xu\*, Wei Wei\*, Yuhan Wu, Lemeng Liang, Shuqi Cai, Zuozheng Liu, Rujing Wang, Mengchao Li*

*Email: taoxu2011@tju.edu.cn*

*Key Laboratory of Smart Grid of Ministry of Education, Tianjin University, Tianjin, China*

**ABSTRACT**

**The international mega-event, such as the Winter Olympic Game, has been considered as one of the most carbon intensive activities worldwide. The commitment of fully renewable energy accommodation and utilization while ensuring the extreme high reliability has brought significant challenges on system operation due to the stochastic nature of the renewables. The battery energy storage system (BESS) composed of stationary energy storage system (SESS) and shared mobile energy storage system (MESS) can be utilized to meet the requirements of short-term load surges, renewable accommodation and emergency power supply for important loads during the mega-event. The BESS can continue to serve the venues' electricity consumption to satisfy the carbon neutrality after the event. On the other hand, the low ambient temperature of Winter Olympic game has significant impact on BESS's degradation and performance which need to be integrated to the charging and discharging model of BESS. To this end, a joint two-stage optimal configuration method considering the ambient temperature of SESS and MESS has been developed to support the mega-event carbon neutrality, to reduce redundant BESS capacity allocation and improve the system life cycle cost-benefit. Simulation results have demonstrated the rationality and effectiveness of the collaborative operation of SESS and the MESS under various scenarios.**

## 1. Introduction

The mega-event, such as Olympic Games, international conferences and exhibitions require high reliabilities on energy supply in lighting, audio, transportation, screen, media and data centers etc., consequently have been considered as one of the most carbon intensive activities worldwide. In March 2020, the International Olympic Committee (IOC) declared that from 2030 onwards all Olympic Games will be required to be 'climate positive' which means the events will need to reduce their carbon emissions in line with the Paris Agreement and compensate more than 100% of their remaining emissions [1]. The organization committee of Tokyo Olympic Game has set the target for the first net zero carbon emission mega-event that relies on renewable energy resources. Without any spectators being present in any venues in Tokyo due to the global epidemic, the Olympic Game in Tokyo has offered a benchmark for mitigating the environmental impact of hosting mega-events with leading-edge technologies, including hydrogen, self-driving electric vehicles (EVs), recycling techniques, as well as the artificial intelligence (AI). In January 2021, Paris 2024 organizing committee pledged to deliver the world's first climate positive Olympic and Paralympic Games [2].

The upcoming Beijing Winter Olympic Game will attempt to be the first carbon-neutral Winter Olympics, aiming to make a real, tangible difference on energy utilization. With 100% renewable power supply to all 26 venues, the carbon emission reduction during the mega-event can be approximately 320,000 tons. After the game, the infrastructure such as REGs, flexible DC grid, energy storage system (ESS) and EV charging facilities etc. will continue to serve the capital's immense electricity consumption to satisfy the carbon neutrality [2].

However, the commitment of fully REG accommodation and utilization while ensuring the operation reliability has brought significant challenges due to the stochastic nature of the renewables, in particular under the extreme weather conditions for winter games [3]-[5]. Therefore, battery energy storage systems (BESSs) have been widely deployed during this mega-event to achieve the balance of carbon footprint reduction and secure energy supply, to meet the challenges from the variability and intermittency of the power generation from non-dispatchable REGs [6]. The BESS can also smooth the fluctuations, create energy arbitrage, improve power quality, participate auxiliary services etc. after the event [7]-[9]. To this end, the efficient BESS utilization can be recognized as one of the key technologies that enable the carbon neutral mega-event.

According to the literature, the redundant configuration and associate cost of BESS are still the major issue that hindered the wide spread integration of BESS. Taking the operation strategies into the consideration of BESS planning, the contradiction of REG accommodation and cost control can be solved to a certain extent. Aiming at the wind power curtailment minimization and economic operation under normal operation, two-stage BESS optimal configuration approaches have been conducted [10]-[11]. By integrating the photovoltaic (PV) and the BESS into schedulable power supply, energy losses can be reduced and voltage stability can be improved [12], the system reliability, and the infrastructure upgrade deferral can be realized through load management under various pricing schemes [13]. For abnormal operation, the sitting and sizing of BESS considering the impact of grid reconfiguration have been determined via iterative solution where the decisions are made based on the minimization of line congestion and voltage deviations simultaneously [14], or through the incorporated utilization of the soft open points (SOPs) and reactive power contribution of distributed generation [15].

The shared BESS means that the ownership and the utilization are different at most of the time. The owner can lease the assets for certain time of a periodic payment to satisfy the requirements of renters who cannot afford the high capital investment of BESSs [16]. In order to maximize the revenue of the asset owner and distribution network operators (DNOs), the shared BESS temporal and spatial operation schemes have been determined for distribution systems with proliferation of renewables [17]-[18]. A distributed optimization strategy of shared BESS based on game theory has been proposed, which aims to reduce the total energy operating cost and the peak-to-average power ratio for the entire grid through the energy capacity trading [19]. For short term trading, taking the charge and discharge order, load profile, pricing and status of available shared BESS into consideration, day-ahead scheduling and rolling can be determined via a cloud decision making strategy that can meet the load requirements at a lower price, and maintain revenue for the owner at the same time [20].

The traditional stationary energy storage system (SESS) has challenged by the relative high cost, low utilization rate and short life. On the other hand, the flexible demand and EV fleet V2G responses are experiencing difficulties on aggregation and time delay which cannot give full play on providing effective operational and emergency services due to the uncertainties on capacity. Consequently, the integrated container size mobile energy storage system (MESS) has attracted widespread attention with configurable capacities, flexible applications, compact and safe design, as well as strong reliability [21]-[23]. Most of the existing research and implementations of MESS focus on the emergency support and fault recovery [24]-[26]. In order to realize a prompt fault recovery, s two-stage scheduling framework with pre-positioning and real-time allocation for MESS has been developed [24], a real time island division strategy aiming at maximizing the recovery of critical loads has been developed, and the MESS optimal configuration method has been provided to support individual island [25]. The MESS can also be utilized to meet the requirement on REG accommodation [27], voltage regulation, loss reduction [28], and uninterrupted power supply (UPS) [29], etc. Apparently, the MESS can be considered as an effective alternative solution for SESS, the cooperative solution can further strengthen the system reliability, improve the renewable integration, as well as maximize the benefit under various scenarios, therefore the joint optimization solution can be implemented for carbon neutral mega-event.

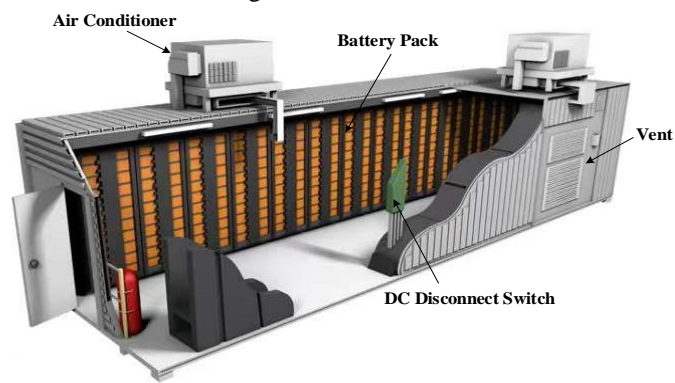

*Fig. 1   Schematic diagram of a typical container BESS*

Most of the BESSs take the container as the carrier to form container energy storage system (CESS) that integrates lithium-ion battery pack, battery management system (BMS), power conversion system (PCS), thermal management system and fire protection system into a standard container as shown in Fig.1. It features with compact design, relatively large capacity, convenient relocation and installation [30]. In this paper, MESS integrates the battery pack and PCS into one container for flexible deployment, while the battery units and PCS of SESS are placed in different containers to facilitate operation and maintenance. However, the confined space of the containers results in excessive temperature rise during BESS operation, and the ambient temperature has a significant impact on the service performance and life of the battery, the reduction of ionic conductivity, the increase of charge-transfer resistance and the emergence of lithium-ion plating cause the degradation on battery capacity and power during low temperature [31]. High temperature conditions also accelerate the aging and shorten the service life of the battery. With the stimulation of temperature rise, exothermic reaction can be triggered which may lead to thermal runaway [32]. Therefore, it is necessary to consider the impact of ambient temperature on lifetime and cost of BESS into the planning stage, in particular for Winter Olympic games with extreme weather conditions. The thermal management of BESS can improve its lifetime and performance, so as to improve the economy and safety of the overall system. The existing literature studies the thermal management system of EV battery [33], battery heating generation monitoring [34]-[36], and thermal management control of BESS [37]. However, the influence of temperature hasn't been studied in-depth at the planning stage, especially for CESS.

To this end, the findings on the joint operation of SESS and MESS have substantial advantages in meeting the strict requirements of carbon neutral mega-event with diversified generation/demand and high reliability on important loads under extreme ambient temperature, the key contributions can be highlighted as follows:

1) A CESS thermal management model which considers the effects of ambient temperature and battery self-heating release on the thermal equilibrium inside the container, and adopts air conditioning and natural ventilation for temperature adjustment has been developed.

2) The life cycle model of BESS is further enriched, taking the self-loss cost of battery and the thermal management cost of CESS into consideration.

3) An operation strategy of SESS is proposed to stabilize the fluctuations caused by high REG permeability and heavy loading conditions under multiple scenarios. The operation strategy determines the active/reactive power of the SESSs and VSCs to maintain the safety and economic operation of the AC/DC hybrid system as a complete one. At the same time, the linearized battery life cycle model and thermal management model are used to support the economic operation of the system.

4) A coordinated operation strategy of SESS and MESS is proposed to ensure the safety and economic operation of the system under heavy loading conditions and the emergencies for mega-event scenarios.

5) A two-stage optimization model of SESS and MESS capacity allocation of AC/DC hybrid system is established. The model reduces the configuration cost of SESS in the project cycle by configuring MESS under special scenarios, ensures the safety and economy of the system through the coordinated operation of SESS and MESS, and can provide emergency power backup for important loads for mega-event.

The remainder of this paper is organized as follows: the external characteristic model, cost-benefit model, thermal management model and life cycle model of BESS are introduced in Section 2. Section 3 presents the two-stage joint optimization of SESS and MESS with the active/reactive coordinated operation strategy of BESS and VSC considering thermal management. Case studies are carried out in Section 4, and the relevant concluding remarks are delivered in Section 5.

## 2. Energy storage system model

### 2.1 Refined BESS characteristic model

The BESS mainly consists of PCS and battery units. The PCS permits the BESS to generate both active and reactive power in all four quadrants when BESS is connected with AC distribution network [38]. The charging and discharging power constraints are shown in (1)-(3), while only (2) is required for DC distribution system.

$$\sqrt{(P_{dis,i,\omega}^{BESS}(t)+P_{ch,i,\omega}^{BESS}(t))^2 + Q_{i,\omega}^{BESS}(t)^2} \leq S_i^{BESS} \tag{1}$$

$$\begin{cases} 0 \leq P_{dis,i,\omega}^{BESS}(t) \leq P_{rate,i}^{BESS}\mu_{dis,i,\omega}(t) \\ -P_{rate,i}^{BESS}\mu_{ch,i,\omega}(t) \leq P_{ch,i,\omega}^{BESS}(t) \leq 0 \\ \mu_{dis,i,\omega}(t)+\mu_{ch,i,\omega}(t)=1 \end{cases} \tag{2}$$

$$-Q_{rate,i}^{BESS} \leq Q_{i,\omega}^{BESS}(t) \leq Q_{rate,i}^{BESS} \tag{3}$$

where, $P_{ch,i,\omega}^{BESS}(t)$ and $P_{dis,i,\omega}^{BESS}(t)$ are the charging/discharging active power of the $i$-th BESS at time $t$ under scenario ω respectively, $Q_{i,\omega}^{BESS}(t)$ is the charging/discharging reactive power of the $i$-th BESS at time $t$ under scenario ω. And the output power of BESS to the grid is set as positive. $S_i^{BESS}$, $P_{rate,i}^{BESS}$ and $Q_{rate,i}^{BESS}$ represent the PCS capacity, maximum active and reactive power of the $i$-th BESS, respectively. BESS can only be in the state of discharging or charging at the same time. So, 0-1 variables, $\mu_{dis,i,\omega}(t)$ and $\mu_{ch,i,\omega}(t)$, representing the discharging/charging sign of the $i$-th BESS at time $t$ under scenario ω, are introduced respectively.

If two sets of BESSs are installed at the same node, the operation constraint (1) and (3) are modified to (4), and (5) respectively. The following constraints are set to ensure that two sets of BESSs absorb or release reactive power at the same time.

$$\sqrt{(P_{dis,i,\omega}^{BESS}(t)+P_{ch,i,\omega}^{BESS}(t))^2 + (Q_{dis,i,\omega}^{BESS}(t)+Q_{ch,i,\omega}^{BESS}(t))^2} \leq S_i^{BESS} \tag{4}$$

$$\begin{cases} 0 \leq Q_{dis,i,\omega}^{BESS}(t) \leq Q_{rate,i}^{BESS} \mu_{dis,i,\omega}(t) \\ -Q_{rate,i}^{BESS} \mu_{ch,i,\omega}(t) \leq Q_{ch,i,\omega}^{BESS}(t) \leq 0 \\ \mu_{dis,i,\omega}(t) + \mu_{ch,i,\omega}(t) = 1 \end{cases} \tag{5}$$

where $Q_{ch,i,\omega}^{BESS}(t)$ and $Q_{dis,i,\omega}^{BESS}(t)$ are the charging/discharging reactive power of the *i*-th BESS at time *t* under scenario ω respectively. Reactive power constraints of BESS installed at the same node use the same 0-1 variables, $\mu_{dis,i,\omega}(t)$ and $\mu_{ch,i,\omega}(t)$.

The state-of-charge (SOC) of BESS describes the chronological relationship between charging/discharging power and SOC of the BESS which is limited by the upper and lower limits in (6) and (7). Moreover, the SOC levels of BESS shall be restored daily to ensure its sustainable operation, as enforced on (8).

$$SOC_{i,\omega}(t) = SOC_{i,\omega}(t-1) \cdot (1-\delta) - \frac{P_{ch,i,\omega}^{BESS}(t)\eta_C \eta_{PCS} \Delta t}{E_{rate,i}} - \frac{P_{dis,i,\omega}^{BESS}(t)\Delta t}{E_{rate,i}\eta_D \eta_{PCS}} \tag{6}$$

$$SOC_{min,i} \leq SOC_{i,\omega}(t) \leq SOC_{max,i} \tag{7}$$

$$SOC_{i,\omega}(0) = SOC_{i,\omega}(T) \tag{8}$$

where $SOC_{i,\omega}(t)$ is the SOC of the *i*-th BESS at time *t* under scenario ω. $SOC_{max,i}$ and $SOC_{min,i}$ represent the threshold of *SOC* of the *i*-th BESS. $E_{rate,i}$ is the rated capacity of the *i*-th BESS and δ is the self-discharging rate of the battery. $\eta_C$ and $\eta_D$ are the charging and discharging efficiencies of the battery and $\eta_{PCS}$ represents the associate PCS efficiency. where $SOC_{i,\omega}(0)$ and $SOC_{i,\omega}(T)$ are the initial and ending SOC of the *i*-th BESS under scenario ω of each operation cycle.

*2.2 Cost-benefit model*

*1) Benefit model*

The BESS invested and managed by the event organizing committee usually intends to maximize the benefits of the venue and associate grid. The interests of investors include not only the direct profits from providing various services to the grid, but the indirect benefits of the power grid, including but not limited to PV consumption, peak shaving and emission reduction *etc*. This paper only quantifies the direct income including arbitrage revenue and network loss reduction to avoid redundant capacity configuration of BESS. The arbitrage revenue of BESS can be formulated as (9) with time-of-use (TOU) pricing scheme:

$$B_{arb} = \sum_{\omega \in W}[D_\omega \sum_{t=1}^{24} \sum_{i=1}^{N_{BESS}} (P_{dis,i,\omega}^{BESS}(t)+P_{ch,i,\omega}^{BESS}(t)) \times \text{price}(t)] \tag{9}$$

where, $B_{arb}$ is the arbitrage revenue of the BESS and the W is the collection of scenarios, $D_\omega$ is the number of days under scenario *ω*. $N_{BESS}$ is the number of sitting locations of BESSs and price(t) is the electricity price at time *t*.

Distributed BESS can effectively reduce network losses and obtain benefits, as shown below:

$$B_{loss} = C_{loss0} - C_{loss} \tag{10}$$

$$C_{loss} = \sum_{\omega \in W} D_\omega \begin{bmatrix} \sum_{t=1}^{24}\sum_{i=1}^{N_{ac}}\sum_{j\in\Omega_i} \frac{1}{2}(R_{ij,ac} \times I_{ij,ac,\omega}^2(t) \times price(t)) \\ +\sum_{t=1}^{24}\sum_{i=1}^{N_{dc}}\sum_{j\in\Omega_i} \frac{1}{2}(R_{ij,dc} \times I_{ij,dc,\omega}^2(t) \times price(t)) \\ +\sum_{t=1}^{24}\sum_{h=1}^{H} (\eta_{loss}^{VSC} \times |P_{ac,h,\omega}^{VSC}(t)| \times price(t)) \end{bmatrix} \tag{11}$$

where $C_{loss0}$ and $C_{loss}$ are the network loss cost before and after BESS participation, respectively. In (11), the first and second items represent the AC and DC line losses, and the third item is the power losses in VSC. $N_{ac}$ and

$N_{dc}$ are the number of nodes in AC and DC subsystem respectively. $R_{ij,ac}$ and $R_{ij,dc}$ are the resistance of the AC and DC branch ij, respectively. $I_{ij,ac,\omega}(t)$ and $I_{ij,dc,\omega}(t)$ are the current flowing from AC node $i$ to $j$ and current flowing from DC node $i$ to $j$, separately. H is the number of VSCs and $\Omega_i$ is the set of adjacent nodes of node $i$. $\eta_{loss}^{VSC}$ is the power losses coefficient of the VSC. $P_{ac,h,\omega}^{VSC}(t)$ is the injected power into the AC system via the $h$-th converter at time $t$ under scenario $\omega$.

*2) Cost model of SESS*

In this paper, the service cycle of SESS is set as the project cycle, the whole life cycle cost (LCC) contains the capital investment, replacement cost, fixed operation and maintenance costs, variable operation and maintenance costs and disposal cost. The capital investment includes the costs of storage unit, the PCS, and the necessary supporting facilities, so it can be expressed as:

$$C_{cap} = \left(c_E E_{rate,i} + c_P P_{rate,i}^{SESS} + c_B E_{rate,i}\right) \frac{\tau(1+\tau)^Y}{(1+\tau)^Y - 1} \quad (12)$$

where $c_E$ is the price of SESS capacity per kW·h, $c_P$ is the PCS cost per kW and $c_B$ is the cost of the necessary supporting facilities per kW·h. Y is the project cycle (year) and $\tau$ is the discount rate (%). $P_{rate,i}^{SESS}$ is the rated power of the SESS.

The lifetime of the lithium-ion battery and the PCS cannot fulfil the requirements of the whole project cycle without replacement. Annual replacement cost can be formulated as:

$$C_{rep} = c_E E_{rate,i} \sum_{\varepsilon=1}^{k} \left( \frac{(1-\alpha)^{\varepsilon n}}{(1+\tau)^{\varepsilon n}} \frac{\tau(1+\tau)^Y}{(1+\tau)^Y - 1} \right) + c_P P_{rate,i}^{SESS} \frac{(1-\alpha)^{10}}{(1+\tau)^{10}} \frac{\tau(1+\tau)^Y}{(1+\tau)^Y - 1} \quad (13)$$

where α is the average annual decline rate of the SESS capital investment and $k$ is the total number of times of battery replacement, $n$ is the lifetime of the chosen SESS, $\varepsilon$ is the sequence of replacement. $n$ can be obtained from the life cycle model, and the lifetime of the PCS is chosen as 10 years in this study.

The fixed operation and maintenance costs include the labour and management cost, related to the types of battery and associate rated power.

$$C_{fix} = c_f P_{rate,i}^{SESS} \quad (14)$$

where $c_f$ is the fixed operation and maintenance cost per kW.

The variable operation and maintenance costs include the battery self loss cost and air conditioning operation fees which changes according to the working conditions and ambient temperature.

$$C_{var} = \sum_{\omega \in W} [D_\omega \sum_{t=1}^{24} \sum_{i=1}^{N_{SESS}} [(P_{hot,i,\omega}^{Air}(t) + P_{cool,i,\omega}^{Air}(t)) \cdot price(t) + I_{bat,i,\omega}^2(t) \cdot R_{int}^{bat} \cdot price(t) \cdot N_{bar}] \cdot N_{CESS,i}] \quad (15)$$

$$I_{bat,i,\omega}(t) = \frac{P_{ch,i,\omega}^{SESS}(t)\eta_C \eta_{PCS} - \dfrac{P_{dis,i,\omega}^{SESS}(t)}{\eta_D \eta_{PCS}}}{N_{par} \cdot U_{bar} \cdot N_{CESS,i}} \quad (16)$$

where $P_{hot,i,\omega}^{Air}(t)$ and $P_{cool,i,\omega}^{Air}(t)$ are the heating and cooling power of each CESS air conditioner of the $i$-th SESS at time $t$ under scenario $\omega$ respectively. $I_{bat,i,\omega}(t)$ and $R_{int}^{bat}$ are the current and internal resistance of single battery unit of CESS. $N_{bar}$ is the number of single cells constituting CESS of each container and $N_{CESS,i}$ is the number of CESS constituting SESS. $N_{par}$ is the number of parallel branches of battery in each CESS.

The treatment cost of the decommissioned battery materials (hazardous and non-hazardous) can be expressed as:

$$C_{dis} = c_{d} P_{rate,i}^{SESS} \sum_{\varepsilon=1}^{k} \left( \frac{(1-\alpha)^{\varepsilon n}}{(1+\tau)^{\varepsilon n}} \frac{\tau(1+\tau)^{Y}}{(1+\tau)^{Y}-1} \right) \quad (17)$$

where $c_d$ is the specific disposal cost per kW of the battery.

*3) Cost model of MESS*

The service cycle of MESS is a special short-term period of time. The operator pays the rent to MESS owner according to capacity and life damage compensation. Lease capacity refers to the energy storage capacity (including energy capacity and rated power) that can meet the needs of the operator to realize the charging and discharging during the dispatching cycle, which directly reflects the BESS demand of the operator. In addition, frequent charging and discharging can cause the aging acceleration of battery units, the life damage compensation can be used to quantify the corresponding degradation cost of MESS. In addition, the operator shall also bear the fixed/variable operation and maintenance cost during the operation of MESS, as shown in (14) to (16). The MESS module adopts the fixed capacity and power matching mode, and is charged according to the number of modules. The MESS rent cost is:

$$C_{rent} = \sum_{i=1}^{N_{MESS}} [c_{rent} N_{MESS,i} T_{rent} + T_{rent} \sum_{j=1}^{N_{MESS,i}} (\frac{\xi_{lin,j}}{0.2} c_{E} E_{rate}^{MESS})] \quad (18)$$

$$\begin{cases} \xi_{lin,j} = \xi_{lin,j}^{idl} + 0.5 \xi_{lin,j}^{cyc} \\ \xi_{lin,j}^{idl} = 1.952 e^{-5} \cdot SOC_{avg,j} + 1.85 e^{-5} \\ \xi_{lin,j}^{cyc} = \sum_{t=1}^{24} 4.9 e^{-5} \cdot DOD_{lin,j}(t) + 1.012 e^{-20} \end{cases} \quad (19)$$

where $N_{MESS}$ is the number of nodes with MESS and $N_{MESS,i}$ is the number of MESS containers installed on node i where MESS is installed. $c_{rent}$ is the daily rent of MESS unit. $T_{rent}$ and $\xi_{lin,j}$ are the rent period and the daily life damage of the *j*-th MESS module respectively. $E_{rate}^{MESS}$ is the rated energy capacity of the MESS module. Battery is usually decommissioned when the degrading capacity approaches 20% of the rated capacity value. The life cycle model in [39] is linearized in the interval of 0-1 to obtain (19), which is used to estimate the life damage, and its norm error is 1.2e-20. $\xi_{lin,j}^{idl}$ and $\xi_{lin,j}^{cyc}$ are the amount of capacity degradation in one day and several cycles, respectively. $SOC_{avg,j}$ is the average daily SOC and $DOD_{lin,j}(t)$ is the cyclic depth of discharge (DOD).

*2.3 Thermal management model*

The thermal management model of BESS is mainly composed of battery heat generation model, battery heat transfer model, container heat balance model and associate constraints.

*1) Battery heat generation model*

The heat generation within the lithium-ion battery at normal operation is associated with the charging transfer and chemical reactions during charging and discharging [40]. The heat is generated mainly in the reversible process, active polarization process and ohmic heating process in battery [32]. The heat production of a single CESS is:

$$Q_{CESS,i,\omega}^{heat}(t) = (I_{bat,i,\omega}^{2}(t) R_{int}^{bat} + I_{bat,i,\omega}(t) T_{CESS,i,\omega} \frac{\partial U_{bar}}{\partial T_{CESS,i,\omega}}) \, N_{bar} / 1000 \quad (20)$$

Where, $Q_{CESS,i,\omega}^{heat}(t)$ represents the heat(kW) generated by the normal operation of CESS and $T_{CESS,i,\omega} \cdot (\partial U_{bar} / \partial T_{CESS,i,\omega})$ is the empirical constant which is chosen as 0.0116 [41].

*2) Battery heat transfer model*

The battery heat transfer model mainly considers the heat released into the air and the self-absorption heat of the battery. The heat transfer of a single CESS is as follows:

$$Q_{CESS,i,\omega}^{heat}(t) = Q_{rel,i,\omega}^{heat}(t) + Q_{abs,i,\omega}^{heat}(t) \tag{21}$$

$$Q_{rel,i,\omega}^{heat}(t) = A_{bar} \cdot h_{trans} \cdot N_{bar}(T_{bar,i,w}(t) - T_{CESS,i,w}(t))/1000 \tag{22}$$

$$Q_{abs,i,\omega}^{heat}(t) = C_{Spe}^{bat} \cdot M^{bat} \cdot N_{bar}(T_{bar,i,w}(t) - T_{CESS,i,w}(t))/(3.6 \times 10^6) \tag{23}$$

where, $Q_{rel,i,\omega}^{heat}(t)$ and $Q_{abs,i,\omega}^{heat}(t)$ are the heat released into the air and the self-absorption heat of the battery, respectively. $A_{bar}$ is the heat dissipation area of a single battery and $h_{trans}$ is the heat transfer coefficient. $T_{bar,i,w}(t)$ and $T_{CESS,i,w}(t)$ are the battery surface temperature and the container internal temperature, respectively. $C_{Spe}^{bat}$ is the specific heat capacity of battery and the $M^{bat}$ is the parameter representing the battery quality.

*3) Container heat balance model*

In this paper, the battery surface temperature modification can be decoupled into two stages. The first stage represents the BESS surface temperature modification from previous time slot to the room temperature at the current time. And the second stage illustrates the temperature changes from current room temperature to normal working condition temperature. Apart of the air conditioning that regulates the temperature of the container, the CESS can also conduct natural ventilation and heat exchange through the vent on board. In addition, passive heat exchange with the external environment can be done through the wall. The heat balance of CESS can be expressed as follows:

$$(T_{CESS,i,w}(t) - T_{CESS,i,w}(t-1)) \cdot C_{Spe}^{air} \cdot M^{air}/(3.6 \times 10^6) = \\ COP \cdot P_{hot,i,\omega}^{Air}(t) - EER \cdot P_{cool,i,\omega}^{Air}(t) + Q_{rel,i,\omega}^{heat}(t) + Q_{vent,i,\omega}^{heat}(t) + Q_{wall,i,\omega}^{heat}(t) - Q_{abstem,i,\omega}^{heat}(t) \tag{24}$$

$$Q_{vent,i,\omega}^{heat}(t) = \Delta t \cdot \alpha_{vent} \cdot C_{Spe}^{air} \cdot \rho^{air}(T_{ext,i,w}(t) - T_{CESS,i,w}(t)) \cdot X_{vent,i,w}(t)/(3.6 \times 10^6) \tag{25}$$

$$\alpha_{vent} = \frac{A_{vent}}{2} \times C_{flo} \sqrt{C_{wind} \cdot V_{wind}^2} \tag{26}$$

$$Q_{wall,i,\omega}^{heat}(t) = K_{wall} \cdot A_{wall}(T_{ext,i,w}(t) - T_{CESS,i,w}(t))/1000 \tag{27}$$

$$Q_{abstem,i,\omega}^{heat}(t) = C_{Spe}^{bat} \cdot M^{bat} \cdot N_{bar}(T_{CESS,i,w}(t) - T_{bar,i,w}(t-1))/(3.6 \times 10^6) \tag{28}$$

where $C_{Spe}^{air}$ and $M^{air}$ represent the specific heat and quality of air, respectively. The parameter $M^{air}$ can be calculated by multiplying the volume of the container with the air density. COP and EER are the energy efficiency ratios of air conditioning heating and cooling respectively. $Q_{vent,i,\omega}^{heat}(t)$, $Q_{wall,i,\omega}^{heat}(t)$ and $Q_{abstem,i,\omega}^{heat}(t)$ are the vent heat exchange, wall heat exchange and the first stage heat exchange of battery temperature variation, respectively. $a_{vent}$ is the volume flow rate[42] and $\rho^{air}$ is air density. $T_{ext,i,w}(t)$ is the external ambient temperature and $X_{vent,i,w}(t)$ is a 0-1 variable indicating the operating status of the vent, where 1 indicates open status of the vent, and vice versa. $A_{vent}$ and $C_{flo}$ are the vent area and discharge coefficient respectively. $C_{wind}$ is the wind effect coefficient and $V_{wind}$ is the wind speed. $K_{wall}$ and $A_{wall}$ are the heat capacity and area of CESS wall.

*4) Constraints*

The CESS internal temperature constraint is:

$$\begin{cases} T_{CESS,min} \leq T_{CESS,i,w}(t) \leq T_{CESS,max} \\ T_{CESS,i,w}(0) = T_{CESS,i,w}(T) \end{cases} \tag{29}$$

where $T_{CESS,min}$ and $T_{CESS,max}$ are the lower and upper limits of CESS internal temperature.

The constraints on the relationship between battery surface temperature and the CESS internal temperature is:

$$T_{CESS,i,w}(t) \leq T_{bar,i,w}(t) \tag{30}$$

Air conditioning power constraints are:

$$0 \leq P_{hot,i,\omega}^{Air}(t) \leq P_{\max}^{Air}(t) \cdot X_{air,i,w}(t) \tag{31}$$

$$0 \leq P_{cool,i,\omega}^{Air}(t) \leq P_{\max}^{Air}(t) \cdot (1 - X_{air,i,w}(t)) \tag{32}$$

where $P_{\max}^{Air}(t)$ is the maximum operating power of the air conditioner.

*2.4 Life cycle model*

The linearized battery life model in Section 2.2 can only be used to guide BESS operation or short-term life estimation. In terms of complete LCC, a more accurate model that can track the capacity degradation and aging performance of BESS is required. The battery degradation model considers the combined effect of calendar aging $\xi_{ref,\omega}^{idl}$ and cycle aging $\xi_{ref,\omega}^{cyc}$ [40] and the daily degradation of the battery $\xi_{ref,\omega}$ can be illustrated in (33) taking the influence of temperature into account [43].

$$\begin{cases} \xi_{ref,\omega} = \xi_{ref,\omega}^{idl} + \xi_{ref,\omega}^{cyc} \\ \xi_{ref,\omega}^{idl} = f_{ref,soc}(SOC_{avg,\omega}) \\ \xi_{ref,\omega}^{cyc} = \sum_{i=1}^{N_{DOD}} f_{ref,dod}(DOD_{ref,\omega}(i)) f_{ref,T}(T_{i,\omega,avg}) \\ f_{ref,soc}(SOC_{avg,\omega}) = k_1 SOC_{avg,\omega}^2 + k_2 SOC_{avg,\omega} \\ f_{ref,dod}(DOD_{ref,\omega}(i)) = k_3 DOD_{ref,\omega}^2(i) + k_4 DOD_{ref,\omega}(i) \\ f_{ref,T}(T_{i,\omega,avg}) = \begin{cases} e^{k_5/T_{i,\omega,avg}} / k_6, & 298K \geq T_{i,\omega,avg} \geq 273K \\ e^{k_7/T_{i,\omega,avg}} / k_8, & 333K \geq T_{i,\omega,avg} > 298K \end{cases} \end{cases} \tag{33}$$

where $SOC_{avg,\omega}$ represents the average SOC under scenario ω and $DOD_{ref,\omega}(i)$ is the DOD of the *i*-th charging and discharging cycle under scenario ω. $N_{DOD}$ is the number of charging and discharging cycles in a day, and $T_{i,\omega,avg}$ is the average temperature of the BESS in the *i*-th charging and discharging cycle. The parameter $k_x$ is usually obtained according to experimental observation and the rain flow method is used to calculate the DOD and the corresponding battery temperature [43].

Consequently, the BESS life can be calculated as follows:

$$n = \frac{0.2}{\sum_{\omega \in W}[D_\omega \cdot \xi_{ref,\omega}]} \tag{34}$$

### 3. Two-stage joint optimization model of SESS and MESS

A two-stage joint optimization model of SESS and MESS capacity allocation in AC/DC hybrid system is established. The first stage optimization mainly solves the optimal allocation of SESS in AC/DC hybrid system under long-term scenarios. Based on the configuration results of the first stage model, the second stage optimization mainly solves the problem of optimal configuration of MESS under heavy loading scenarios during mega-event, taking the coordination of SESS and MESS and the power conservation of important loads into account. The framework of the proposed two-stage joint optimization model is illustrated in Fig.2.

The proposed optimal configuration method of BESS aims to reduce the redundant allocation cost of SESS through the utilization of MESS in a short-term perspective, and improve the economy of the overall system. The two-stage joint optimization model is solved iteratively by the inner and the outer level model. The outer level model mainly concentrates on the capacity of the SESS, or the number of the MESS. And the economic operation of the system is settled by the inner level optimization with the SESS/MESS capacity, rated power and initial SOC determined by the outer optimization process. The outer model is solved by genetic algorithm (GA) with simulated annealing (SA), and the inner model is solved by the mixed integer second-order cone programming (MISOCP).

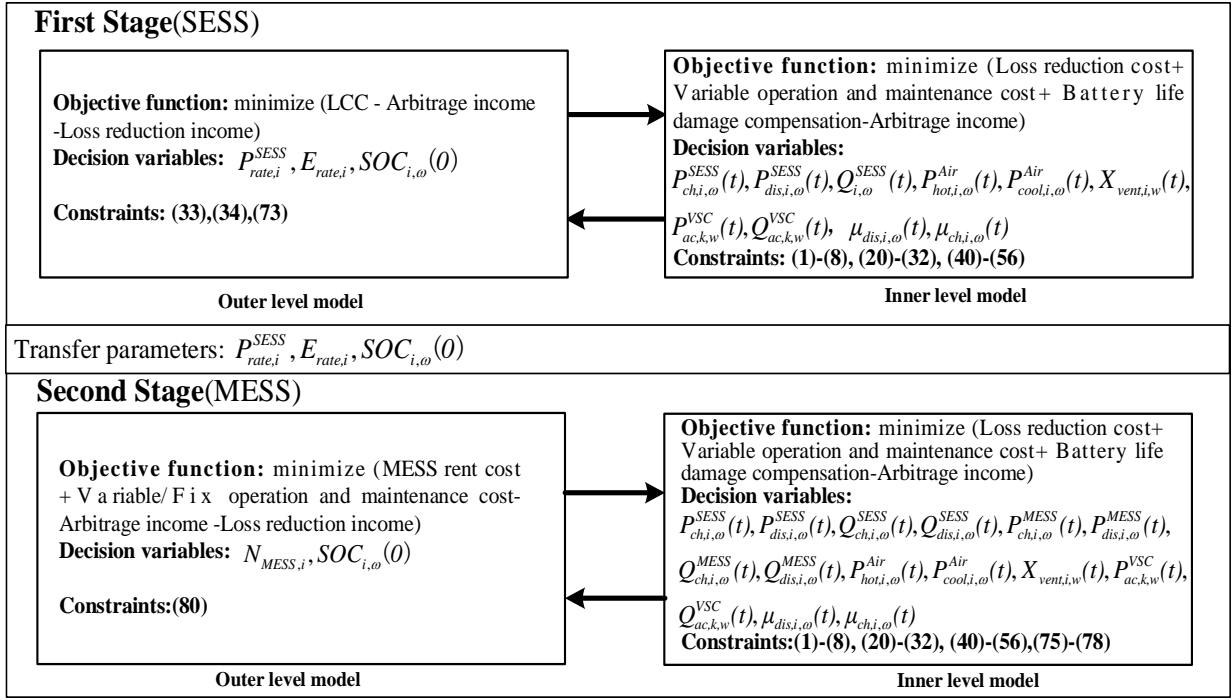

*Fig. 2 Framework of the proposed two-stage joint optimization model of SESS and MESS*

### 3.1 First stage optimization model

1) *Inner level model*

The inner level model is based on the BESS capacity, rated power and the initial SOC from the outer level to optimize the operation of the AC/DC hybrid system under different scenario, considering the cooperative operation of active and reactive power of SESS and voltage source converter (VSC). The linearized battery life model is used to guide the economic operation of SESS, and the influence of thermal management on its economic operation is considered at the same time. The objective function of inner level model under scenario ω is as follows:

$$F_{in,\omega}^{first} = \min(\lambda_1 C_{loss,\omega} + \lambda_2 (C_{var,\omega} + C_{com,\omega} - B_{arb,\omega})) \tag{35}$$

$$C_{loss,\omega} = \begin{bmatrix} \sum_{t=1}^{24}\sum_{i=1}^{N_{ac}}\sum_{j\in\Omega_i} \frac{1}{2}(R_{ij,ac} \times I_{ij,ac,\omega}^2(t) \times price(t)) \\ + \sum_{t=1}^{24}\sum_{i=1}^{N_{dc}}\sum_{j\in\Omega_i} \frac{1}{2}(R_{ij,dc} \times I_{ij,dc,\omega}^2(t) \times price(t)) \\ + \sum_{t=1}^{24}\sum_{h=1}^{H}(\eta_{loss}^{VSC} \times |P_{ac,h,\omega}^{VSC}(t)| \times price(t)) \end{bmatrix} \tag{36}$$

$$C_{var,\omega} = \sum_{t=1}^{24}\sum_{i=1}^{N_{SESS}}[[(P_{hot,i,\omega}^{Air}(t)+P_{cool,i,\omega}^{Air}(t))\cdot price(t)+I_{bat,i,\omega}^2(t)\cdot R_{int}^{bat}\cdot price(t)\cdot N_{bar}]\cdot N_{CESS,i}] \tag{37}$$

$$C_{com,\omega} = \sum_{i=1}^{N_{SESS}}(\frac{\xi_{lin,i,\omega}}{0.2}c_E E_{rate,i}) \tag{38}$$

$$B_{arb,\omega} = \sum_{t=1}^{24} \sum_{i=1}^{N_{SESS}} (P_{dis,i,\omega}^{SESS}(t) + P_{ch,i,\omega}^{SESS}(t)) \times price(t) \tag{39}$$

where $\lambda_1$ and $\lambda_2$ are the weighting coefficients and $C_{com}$ is the compensation cost of battery life expenditure.

Furthermore, in addition to the BESS constraints shown in (1)-(8) and (18)-(32), the inner level model should also meet the AC/DC hybrid system operation constraints. In this study, the main VSC is chosen for $U_{dc}Q$ control, and other are under PQ control. Hence, the AC side nodes of each converter are PQ nodes in the AC subsystem and the DC side nodes are set with constant P. Then the operation constraints are:

(1) Power flow constraint of AC branch

$$\begin{cases} \sum_{l \in \varphi_i} P_{il,ac,\omega}(t) = \sum_{j \in \phi_i} \left( P_{ji,ac,\omega}(t) - R_{ji,ac}(I_{ji,ac,\omega}(t)^2) \right) + P_{i,ac,\omega}(t) \\ \sum_{l \in \varphi_i} Q_{il,ac,\omega}(t) = \sum_{j \in \phi_i} \left( Q_{ji,ac,\omega}(t) - X_{ji,ac}(I_{ji,ac,\omega}(t)^2) \right) + Q_{i,ac,\omega}(t) \end{cases} \tag{40}$$

$$I_{ij,ac,\omega}(t)^2 = \frac{(P_{ij,ac,\omega}(t)^2 + Q_{ij,ac,\omega}(t)^2)}{V_{i,ac,\omega}(t)^2} \tag{41}$$

$$V_{j,ac,\omega}(t)^2 = V_{i,ac,\omega}(t)^2 - 2(R_{ij,ac}P_{ij,ac,\omega}(t) + X_{ij,ac}Q_{ij,ac,\omega}(t)) + (R_{ij,ac}^2 + X_{ij,ac}^2)(I_{ij,ac,\omega}(t)^2) \tag{42}$$

where, $\phi_i$ is the set of receiving end nodes that receive power from node i, and $\varphi_i$ is the set of sending end nodes that have receiving end nodes $i$. $R_{ij,ac}$ and $X_{ij,ac}$ are the resistance and reactance of the AC branch ij, respectively. $P_{ij,ac,\omega}(t)$ and $Q_{ij,ac,\omega}(t)$ are the active and reactive power from AC node $i$ to AC node $j$ at time t, respectively. $I_{ij,ac,\omega}(t)$ and $V_{i,ac,\omega}(t)$ are the current flowing from AC node $i$ to AC node $j$ and the voltage of AC node $i$ at time $t$, respectively. $P_{i,ac,\omega}(t)$ and $Q_{i,ac,\omega}(t)$ are the sum of active and reactive power injected into AC node $i$ at time t, and their expressions are as follows:

$$\begin{cases} P_{i,ac,\omega}(t) = P_{i,ac,\omega}^{PV}(t) + P_{dis,i,ac,\omega}^{SESS}(t) + P_{ch,i,ac,\omega}^{SESS}(t) + P_{i,ac,\omega}^{VSC}(t) - P_{i,ac,\omega}^{LOAD}(t) \\ Q_{i,ac,\omega}(t) = Q_{i,ac,\omega}^{SESS}(t) + Q_{i,ac,\omega}^{VSC}(t) - Q_{i,ac,\omega}^{LOAD}(t) \end{cases} \tag{43}$$

where, $P_{i,ac}^{PV}(t)$ is the active power injected by PV on AC node $i$ at time $t$. $P_{dis,i,ac,\omega}^{SESS}(t)$, $P_{ch,i,ac,\omega}^{SESS}(t)$, $Q_{i,ac,\omega}^{SESS}(t)$, $P_{i,ac,\omega}^{VSC}(t)$, $Q_{i,ac,\omega}^{VSC}(t)$ and $P_{i,ac,\omega}^{LOAD}(t)$, $Q_{i,ac,\omega}^{LOAD}(t)$ are active and reactive power of SESS injected into AC node $i$ during $t$ period, active and reactive power injected into AC side of VSC, and active and reactive power consumed by load during $t$ period. $P_{i,ac,\omega}^{VSC}(t)$ is set as positive from DC subsystem to AC subsystem.

(2) Power flow constraint of DC branch

$$\sum_{l \in \varphi_i} P_{il,dc,\omega}(t) = \sum_{j \in \phi_i} \left( P_{ji,dc,\omega}(t) - R_{ji,dc}(I_{ji,dc,\omega}(t)^2) \right) + P_{i,dc,\omega}(t) \tag{44}$$

$$I_{ij,dc,\omega}(t)^2 = \frac{P_{ij,dc,\omega}(t)^2}{V_{i,dc,\omega}(t)^2} \tag{45}$$

$$V_{j,dc,\omega}(t)^2 = V_{i,dc,\omega}(t)^2 - 2R_{ij,dc}P_{ij,dc,\omega}(t) + R_{ij,dc}^2(I_{ij,dc,\omega}(t)^2) \tag{46}$$

where $R_{ij,dc}$ is the resistance of the DC branch ij. There is no reactive power in DC subsystem and $P_{ij,dc,\omega}(t)$ is the active power from DC node $i$ to DC node $j$ at time $t$. $I_{ij,dc,\omega}(t)$ and $V_{i,dc,\omega}(t)$ are the current flowing from DC node $i$ to $j$ and the voltage of DC node $i$ at time $t$, respectively. $P_{i,dc,\omega}(t)$ is the sum of active power injected into DC node $i$ at time $t$, and can be expressed as:

$$P_{i,dc,\omega}(t) = P_{i,dc,\omega}^{PV}(t) + P_{dis,i,dc,\omega}^{SESS}(t) + P_{ch,i,dc,\omega}^{SESS}(t) - P_{i,dc,\omega}^{VSC}(t) - P_{i,dc,\omega}^{LOAD}(t) \tag{47}$$

where, $P_{i,dc,\omega}^{PV}(t)$ is the active power injected by PV on DC node $i$ at time $t$. $P_{dis,i,dc,\omega}^{SESS}(t)$, $P_{ch,i,dc,\omega}^{SESS}(t)$, $P_{i,ac,\omega}^{VSC}(t)$ and $P_{i,ac,\omega}^{LOAD}(t)$ are active power of SESS injected into DC node $i$, active power flowing through DC side of VSC, and active power consumed by load during $t$ period.

(3) VSC operation power constraints

$$\sqrt{P_{ac,h,\omega}^{VSC}(t)^2 + Q_{ac,h,\omega}^{VSC}(t)^2} \leq S_h^{VSC} \tag{48}$$

$$-P_{ac,max,h}^{VSC} \leq P_{ac,h,\omega}^{VSC}(t) \leq P_{ac,max,h}^{VSC} \tag{49}$$

$$-Q_{ac,max,h}^{VSC} \leq Q_{ac,h,\omega}^{VSC}(t) \leq Q_{ac,max,h}^{VSC} \tag{50}$$

where $S_h^{VSC}$, $P_{ac,max,h}^{VSC}$ and $Q_{ac,max,h}^{VSC}$ are the access capacity of the *h*-th VSC, the upper limit of active power and reactive power transmitted by VSC respectively.

The active power loss of a converter is related to the quantity of the active power and the value of the current flowing through it. As the rated capacity and loading rate of the converter increase, its operating efficiency increases. Reference [44] approximates that the active power losses of the converter is proportional to the active power flowing through the converter. The expressions of active power losses and active power transmitted by VSC are:

$$P_{loss,h,\omega}^{VSC} = \eta_{loss}^{VSC} \cdot \left| P_{ac,h,\omega}^{VSC} \right| \tag{51}$$

$$P_{dc,h,\omega}^{VSC} = P_{ac,h,\omega}^{VSC} - P_{loss,h,\omega}^{VSC} \tag{52}$$

where $\eta_{loss}^{VSC}$ is the active power loss coefficient of the VSC, and its value is generally chosen as 0.03~0.10 [45]. $P_{loss,h,\omega}^{VSC}$ is the active power losses of the *h*-th VSC.

(4) Operating voltage level constraints

$$(V_{i,ac}^{min})^2 \leq (V_{i,ac,\omega}(t))^2 \leq (V_{i,ac}^{max})^2 \tag{53}$$

$$(V_{i,dc}^{min})^2 \leq (V_{i,dc,\omega}(t))^2 \leq (V_{i,dc}^{max})^2 \tag{54}$$

where $V_{i,ac}^{max}$ and $V_{i,ac}^{min}$ are the upper and lower limits of the voltage of the AC node *i*. $V_{i,dc}^{max}$ and $V_{i,dc}^{min}$ are the upper and lower limits of the voltage of the DC node *i*, respectively.

(5) Branch current constraints

$$0 \leq (I_{ij,ac,\omega}(t))^2 \leq (I_{ij,ac}^{max})^2 \tag{55}$$

$$0 \leq (I_{ij,dc,\omega}(t))^2 \leq (I_{ij,dc}^{max})^2 \tag{56}$$

where $I_{ij,ac}^{max}$ is the maximum current of the AC branch ij and $I_{ij,dc}^{max}$ is the maximum current of the DC branch ij.

The operation optimization problem of the AC/DC hybrid distribution system with BESS thermal management is a non-convex nonlinear optimization problem, and the corresponding model can be converted into a cone programming problem by cone transformation in advance for efficient solution.

The second order terms $I_{ij,ac,\omega}^2(t)$, $I_{ij,dc,\omega}^2(t)$ and $I_{bat,i,\omega}^2(t)$ in (36) and (37) can be replaced by $I_{ij,ac,\omega,2}(t)$, $I_{ij,dc,\omega,2}(t)$ and $I_{bat,i,\omega,2}(t)$ for additional relaxation constraints shown in (57). The equation (36) contains an absolute value term $\left|P_{ac,h,\omega}^{VSC}(t)\right|$, the auxiliary variable $P_{h,\omega}^{VSC}(t) = \left|P_{ac,h,\omega}^{VSC}(t)\right|$ is introduced and the constraints are added as show in (58). Equation (51) takes the same conversion form. Then objective function is linearized as (59) and (60).

$$I_{bat,i,\omega,2}(t) \geq I_{bat,i,\omega}^2(t) \tag{57}$$

$$\begin{cases} P_h^{VSC}(t) \geq 0 \\ P_h^{VSC}(t) \geq P_{ac,h}^{VSC}(t) \\ P_h^{VSC}(t) \geq -P_{ac,h}^{VSC}(t) \end{cases} \tag{58}$$

$$C_{loss,\omega} = \begin{bmatrix} \sum_{t=1}^{24} \sum_{i=1}^{N_{ac}} \sum_{j \in \Omega_i} \frac{1}{2} (R_{ij,ac} \times I_{ij,ac,\omega}^2(t) \times price(t)) \\ + \sum_{t=1}^{24} \sum_{i=1}^{N_{dc}} \sum_{j \in \Omega_i} \frac{1}{2} (R_{ij,dc} \times I_{ij,dc,\omega}^2(t) \times price(t)) \\ + \sum_{t=1}^{24} \sum_{h=1}^{H} (\eta_{loss}^{VSC} \times P_{h,\omega}^{VSC}(t) \times price(t)) \end{bmatrix} \tag{59}$$

$$C_{\text{var},\omega} = \sum_{t=1}^{24} \sum_{i=1}^{N_{\text{BESS}}} [(P_{hot,i,\omega}^{Air}(t) + P_{cool,i,\omega}^{Air}(t)) \cdot price(t) + I_{bat,i,\omega,2}(t) \cdot R_{int}^{bat} \cdot price(t) \cdot N_{\text{bar}}] \cdot N_{CESS,i} \quad (60)$$

After the above conversion, the objective function becomes a linear function of the decision variable.

The operation constraints of the system are also linearized by replacing the second term. For AC distribution system operation constraints, the quadratic terms $I_{ij,ac,\omega}^2(t)$ and $V_{i,ac,\omega}^2(t)$ in equations (40)-(43) and equations (53) and (55) are replaced by $I_{ij,ac,\omega,2}(t)$ and $V_{i,ac,\omega,2}(t)$, as shown in (61)-(65). The constraint conditions (62) are further relaxed, as shown in (66), and then equivalently transformed into the standard second-order cone form shown in (67). And the operation constraints of DC distribution system take the same transformation form.

$$\begin{cases} \sum_{l \in \varphi i} P_{il,ac,\omega}(t) = \sum_{j \in \phi i} \left( P_{ji,ac,\omega}(t) - R_{ji,ac}(I_{ji,ac,\omega,2}(t)) \right) + P_{i,ac,\omega}(t) \\ \sum_{l \in \varphi i} Q_{il,ac,\omega}(t) = \sum_{j \in \phi i} \left( Q_{ji,ac,\omega}(t) - X_{ji,ac}(I_{ji,ac,\omega,2}(t)) \right) + Q_{i,ac,\omega}(t) \end{cases} \quad (61)$$

$$I_{ij,ac,\omega,2}(t) = \frac{(P_{ij,ac,\omega}(t)^2 + Q_{ij,ac,\omega}(t)^2)}{V_{i,ac,\omega,2}(t)} \quad (62)$$

$$V_{j,ac,\omega,2}(t) = V_{i,ac,\omega,2}(t) - 2(R_{ij,ac}P_{ij,ac,\omega}(t) + X_{ij,ac}Q_{ij,ac,\omega}(t)) + (R_{ij,ac}^2 + X_{ij,ac}^2)(I_{ij,ac,\omega,2}(t)) \quad (63)$$

$$(V_{i,ac}^{\min})^2 \leq V_{i,ac,\omega,2}(t) \leq (V_{i,ac}^{\max})^2 \quad (64)$$

$$0 \leq I_{ij,ac,\omega,2}(t) \leq (I_{ij,ac}^{\max})^2 \quad (65)$$

$$I_{ij,ac,\omega,2}(t) \geq \frac{(P_{ij,ac,\omega}(t)^2 + Q_{ij,ac,\omega}(t)^2)}{V_{i,ac,\omega,2}(t)} \quad (66)$$

$$\left\| \begin{array}{c} 2P_{ij,ac,\omega}(t) \\ 2Q_{ij,ac,\omega}(t) \\ I_{ij,ac,\omega,2}(t) - V_{i,ac,\omega,2}(t) \end{array} \right\|_2 \leq I_{ij,ac,\omega,2}(t) + V_{i,ac,\omega,2}(t) \quad (67)$$

The charging and discharging constraints shown in (48) of the VSC is a non-linear constraint which can be equivalently transformed into the rotation cone constraint as (68).

$$P_{h,\omega}^{VSC}(t) \times P_{h,\omega}^{VSC}(t) + Q_{k,\omega}^{VSC}(t) \times Q_{k,\omega}^{VSC}(t) \leq 2 \frac{S_h^{VSC}}{\sqrt{2}} \times \frac{S_h^{VSC}}{\sqrt{2}} \quad (68)$$

For constraint (1) in the BESS model, the same transformation is adopted as (69).

$$(P_{dis,i}^{BESS}(t) + P_{ch,i,\omega}^{BESS}(t)) \times (P_{dis,i,\omega}^{BESS}(t) + P_{ch,i,\omega}^{BESS}(t)) + Q_{i,\omega}^{BESS}(t) \times Q_{i,\omega}^{BESS}(t) \leq 2 \frac{S_i^{BESS}}{\sqrt{2}} \times \frac{S_i^{BESS}}{\sqrt{2}} \quad (69)$$

There are multiplicative variables $(T_{CESS,i,w}(t)) \cdot X_{vent,i,w}(t))$ in BESS thermal management constraint (25), so constraint conversion can be carried out. The auxiliary variable $T_{CESS,i,w}^X(t) = T_{CESS,i,w}(t)) \cdot X_{vent,i,w}(t)$ is introduced and the constraints are added as show in (70).

$$\begin{cases} Q_{vent,i,\omega}^{heat}(t) = \Delta t \times \alpha_{vent} \times C_{Spe}^{air} \times \rho^{air}(T_{ext,i,w}(t) \cdot X_{vent,i,w}(t) - T_{CESS,i,w}^X(t))/(3.6 \times 10^6) \\ T_{CESS,i,w}^X(t) \leq T_{CESS,i,w}(t) \\ T_{CESS,i,w}^X(t) \geq T_{CESS,i,w}(t) - T_{CESS,\max} \cdot (1 - X_{vent,i,w}(t)) \\ T_{CESS,\min} \leq T_{CESS,i,w}^X(t) \leq T_{CESS,\max} \end{cases} \quad (70)$$

After the above steps, the cone transformation of the optimization model is completed, namely, the objective function is linearized, and the nonlinear constraint is transformed into linear constraint, second-order cone constraint or rotating cone constraint.

Reference [46] has proved that the second-order conic relaxation (SOCR) is accurate when the objective function is a strictly increasing function of branch current for power flow optimization of distribution systems. The infinite norm of the SOCR deviation is defined as shown in equation (71) to verify the accuracy of the relaxed formula (66) at the optimal solution [38]. When the relaxation deviation is small enough, the SOCR is considered to meet the requirements of calculation accuracy.

$$gap = \left\| I_{ij}(t) - \frac{(P_{ij}(t))^2 + (Q_{ij}(t))^2}{V_i(t)} \right\|_\infty \quad (71)$$

The operation optimization model of the AC/DC hybrid system connected to high penetrated PV and BESS is a large-scale nonlinear optimization problem. The SOCP is a mathematical programming on a convex cone in a linear space. It has the characteristics of fast solution speed and strong optimization capability, so it is suitable for solving large-scale nonlinear optimization problems such as operation sequence optimization of AC/DC hybrid system connected with BESS. In the MATLAB environment, yalmip and Gurobi tools are used to solve the operation optimization model in this study.

2) *Outer level model*

The first stage model optimizes the long-term multi scenario SESS configuration, So the objective of the outer level model is to minimize the total annual cost of SESS, which is the LCC of the SESS minus the average annual network loss reduction and arbitrage income.

$$F_{out}^{first} = \min\begin{pmatrix} C_{cap} + C_{rep} + C_{fix} + C_{var} + C_{dis} - B_{arb} - B_{loss} & \text{if } gap < gap_{max} \\ C_{cap} + C_{rep} + C_{fix} + C_{var} + C_{dis} - B_{arb} - B_{loss} + C_{pun} & \text{if } gap \geq gap_{max} \end{pmatrix} \quad (72)$$

where $C_{pun}$ is the cost penalty applied by the utility for unsatisfied relaxation deviation issues and $gap_{max}$ is the max relaxation deviation.

The optimized variables are the rated capacity, rated power, and initial SOC value of the SESS at each location, as show in Fig.2. The above variable should satisfy the following constraints:

$$\begin{cases} P_{SESS}^{min} \leq P_{rate,i}^{SESS} \leq P_{SESS}^{max} \\ E_{SESS}^{min} \leq E_{rate,i} \leq E_{SESS}^{max} \\ SOC_{min} \leq SOC_{i,\omega}(0) \leq SOC_{max} \\ \sum_{i=1}^{N_{SESS}} c_E E_{rate,i} + c_P P_{rate,i}^{SESS} + c_B E_{rate,i} \leq C_{bud} \end{cases} \quad (73)$$

where $P_{SESS}^{min}$ and $P_{SESS}^{max}$ are the minimum and maximum configured power of SESS. $E_{SESS}^{min}$ and $E_{SESS}^{max}$ are the minimum and maximum investment capacity of the SESS. $P_{rate,i}^{SESS}$ and $E_{rate,i}$ are the rated active power and rated capacity of SESS, respectively. $C_{bud}$ is the initial investment budget of the project.

The upper optimization model is solved by GA combined with SA and the principle of the algorithm is the same as that shown in reference [47]. When GA is used, four genes are used to encode the initial value of each SESS capacity, rated power and initial SOC, and another gene is added as the indicator of PCS reactive power operation. When the gene bit is 1, PCS reactive power can be obtained in the configuration scheme, and when the gene bit is 0, PCS reactive power is ignored. The specific flow chart is shown in Fig. 3.

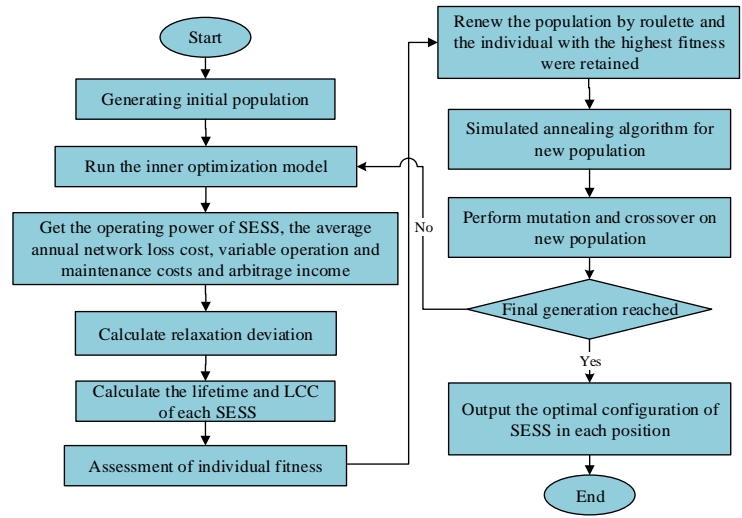

*3.2 Second stage optimization model*

1) *Inner level model*

The inner level model is based on the BESS capacity, rated power and initial SOC of SESS obtained from the first stage optimization and the number of CESSs from the outer level model to optimize the operation of the AC/DC hybrid system under the heavy loading scenario during the event, considering the cooperative operation of active and reactive power of SESS, MESS and VSC. In addition, MESS shall be responsible for ensuring the emergency power supply of important loads. Similar with the first stage optimization model, the linearized BESS life model is used to guide the economic operation of MESS and SESS, and the influence of temperature management on its economic operation is considered at the same time. The objective function of inner level model under this scenario is as follows:

$$F_{in}^{first} = \min(\lambda_1 C_{loss} + \lambda_2 (C_{var,MESS} + C_{com,MESS} - B_{arb,MESS} + C_{var,SESS} + C_{com,SESS} - B_{arb,SESS})) \quad (74)$$

Compared with the first stage optimization model, the AC/DC hybrid system operation constraints (43),(47) are modified to (75),(76).

$$\begin{cases} P_{i,ac,\omega}(t) = P_{i,ac,\omega}^{PV}(t) + P_{dis,i,ac,\omega}^{SESS}(t) + P_{ch,i,ac,\omega}^{SESS}(t) + P_{dis,i,ac,\omega}^{MESS}(t) + P_{ch,i,ac,\omega}^{MESS}(t) + P_{i,ac,\omega}^{VSC}(t) - P_{i,ac,\omega}^{LOAD}(t) \\ Q_{i,ac,\omega}(t) = Q_{dis,i,ac,\omega}^{SESS}(t) + Q_{ch,i,ac,\omega}^{SESS}(t) + Q_{dis,i,ac,\omega}^{MESS}(t) + Q_{ch,i,ac,\omega}^{MESS}(t) + Q_{i,ac,\omega}^{VSC}(t) - Q_{i,ac,\omega}^{LOAD}(t) \end{cases} \quad (75)$$

$$P_{i,dc,\omega}(t) = P_{i,dc,\omega}^{PV}(t) + P_{dis,i,dc,\omega}^{SESS}(t) + P_{ch,i,dc,\omega}^{SESS}(t) + P_{dis,i,dc,\omega}^{MESS}(t) + P_{ch,i,dc,\omega}^{MESS}(t) - P_{i,dc,\omega}^{VSC}(t) - P_{i,dc,\omega}^{LOAD}(t) \quad (76)$$

Since MESS is responsible for the UPS for important loads, its SOC constraints change as follows:

$$\begin{cases} SOC_{min,g}(t) = \dfrac{\mu_{impor,i} \cdot (P_{i,ac}^{LOAD}(t) + P_{i,ac}^{LOAD}(t+1))\Delta t}{E_{rate,i}\eta_D\eta_{PCS}} & \text{if } i \text{ is AC node} \\ SOC_{min,g}(t) = \dfrac{\mu_{impor,i} \cdot (P_{i,dc}^{LOAD}(t) + P_{i,dc}^{LOAD}(t+1))\Delta t}{E_{rate,i}\eta_D\eta_{PCS}} & f \ i \text{ is DC node} \end{cases} \quad (77)$$

$$SOC_{min,g}(t) \leq SOC_{g,\omega}(t) \leq SOC_{max,g} \quad (78)$$

where $\mu_{impor,i}$ is the important load ratio, and $g$ is the MESS installed on the $i$-th node.

In addition, other models and solutions are the same as the first stage inner level optimization model.

3) *Outer level model*

The second stage model optimizes the MESS configuration under short-term heavy loading scenario during the mega-event, So the objective of the outer level model is to minimize the total cost of the MESS and SESS, which includes the rent fees of the MESS and variable operation and maintenance cost of the MESS and SESS minus the network loss reduction and arbitrage revenue.

$$F_{out}^{first} = \min \begin{pmatrix} C_{rent} + C_{var} + C_{fix} - B_{arb} - B_{loss} & \text{if } gap < gap_{max} \\ C_{rent} + C_{var} + C_{fix} - B_{arb} - B_{loss} + C_{pun} & \text{if } gap \geq gap_{max} \end{pmatrix} \quad (79)$$

The optimized variables are the number of MESS unit module and initial SOC of the MESS. as shown in Fig.2. The above variable should satisfy the following constraints:

$$\begin{cases} N_{MESS,i} \leq N_{MESS,max} \\ C_{rent} + C_{var} \leq C_{bud} \end{cases} \quad (80)$$

where $N_{MESS,max}$ is the maximum number of MESS unit.

The upper optimization model is also solved by GA combined with SA as same as the first stage and the flow chart can be found in details in Fig. 4.

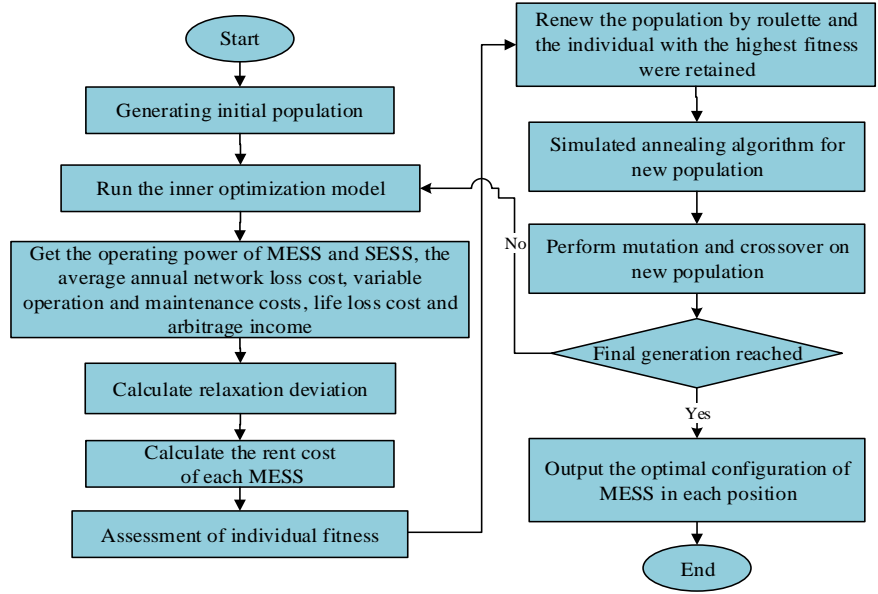

*Fig. 4 Flow chart of the outer optimization model*

## 4. Case studies

### 4.1 Profiles

A simplified medium-voltage AC/DC hybrid distribution system for Winter Olympic venue is used for this study to verify the proposed optimal configuration model considering the joint optimization of SESS and MESS, as shown in Fig. 5. The operation power of the loads and PVs are shown in Appendix Fig.I. The case study network contains 21 nodes and can be divided into three parts, namely, AC system 1, AC system 2, and DC system. In the case study, VSC1 is selected as the balance node of DC subsystem, which adopts $U_{dc}Q$ control and VSC2 adopts PQ control. The voltage level of the DC system is ±10kV. Node 1 on DC side of VSC1 is the balance node of DC subsystem. Node 6 is the balance node of AC distribution system 1 and node 15 is the balance node of AC distribution system 2. The voltage level of both AC distribution systems is 10kV. The upper and lower limits of the voltage magnitude are 1.03 p.u. and 0.97 p.u., respectively. The maximum current-carrying capacity of each branch is 500A. TOU electricity prices are implemented in this area where the peak period (17:00-23:00) is 0.196 \$/(kW·h); the flat period (07:00-17:00) is 0.116 \$/(kW·h); the off-peak period (23:00-07:00) is 0.044 \$/(kW·h). The loading ratios of important nodes (4, 13 and 20) are set as 100%, 40% and 60% respectively. And the wind speed and ambient temperature for each typical scenario are shown in Appendix Fig. II.

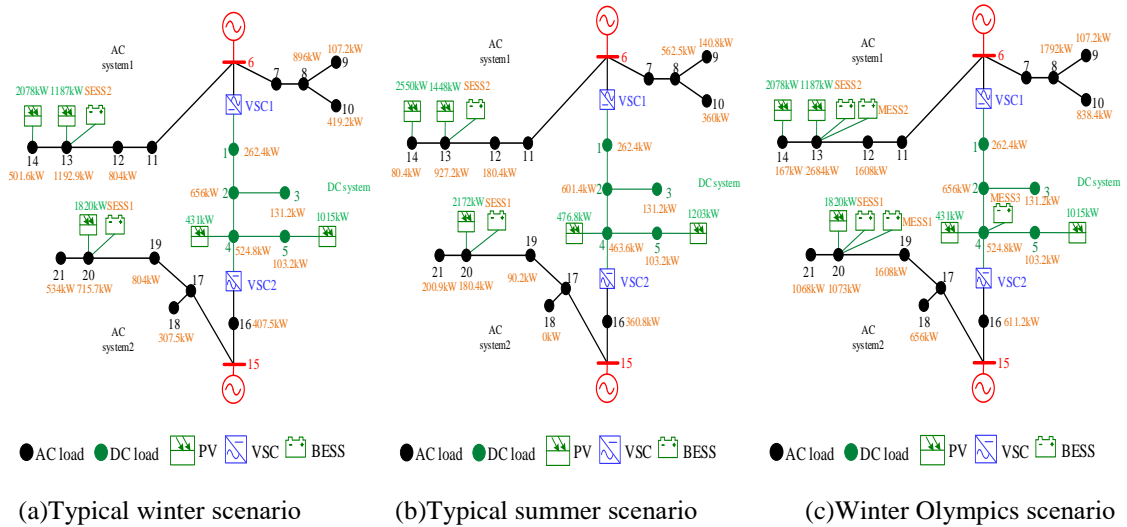

(a)Typical winter scenario  (b)Typical summer scenario  (c)Winter Olympics scenario

*Fig. 5 A simplified medium-voltage AC/DC hybrid distribution system*

The upper and lower limits of the SOC are 90% and 10%, respectively. The project life cycle is set as 20 years without considering the annual average decline rate of the battery-installation cost. The overall duration of the Winter Olympics is approximately 60 days and the parameters of BESS are shown in Table 1[31,41,42,43,47]. The duration of winter scene and summer scene is 6 months respectively. The parameters of VSC are shown in Table 2 and VSC loss coefficient $\eta_{loss}^{VSC}$ is 0.03 in this study. The specification of container for CESS is 12.192m×2.438m×2.896m in size. In the inner model objective function, the weighting coefficient λ1 and λ2 can be obtained by applying the Analytic Hierarchy Process (AHP), λ1=0.67, λ2=0.33.

*Table 1. Initial parameters of BESS model*

| | | | | | | | | | |
|---|---|---|---|---|---|---|---|---|---|
| **External characteristic parameters** | $\eta_C$ | $\eta_D$ | $\eta_{PCS}$ | | | | | | |
| | 0.976 | 0.976 | 0.95 | | | | | | |
| **Economic parameters** | $c_E$ | $c_P$ | $c_B$ | $c_f$ | $c_d$ | $c_{rent}$ | | | |
| | 156$/kW·h | 10$/kW | 0 | 23.8 $/kW·year | 243.4 $/kW | 102.6$/day | | | |
| **Thermal management parameters** | $R_{int}^{bat}$ | $N_{par}$ | $U_{bar}$ | $N_{bar}$ | $A_{bar}$ | h | $C_{spe}^{bat}$ | $M^{bat}$ | $C_{spe}^{air}$ |
| | 0.003Ω | 12 | 0.851kV | 2760 | 0.0418m² | 5w/(m²·K) | 956J/(kg·K) | 1.7kg | 1.003kJ/(kg·K) |
| | $M^{air}$ | COP | EER | $\rho^{air}$ | $K_{wall}$ | $A_{wall}$ | $A_{vent}$ | $C_{flo}C_{wind}^{0.5}$ | |
| | 107.4kg | 3.25 | 3.34 | 1.248kg/m³ | 0.6w/(m²·K) | 114.46m² | 0.5 | 0.29 | |
| **Lifetime parameters** | $k_1$ | $k_2$ | $k_3$ | $k_4$ | $k_5$ | $k_6$ | $k_7$ | $k_8$ | |
| | 6.81e-5 | 4.02e-5 | 3.01e-5 | 8.98e-6 | 6.298e3 | 1.214e10 | -4.665e3 | 1.675e-6 | |

*Table 2. Initial configuration parameters of converters*

| voltage source converter | working mode | Upper limit of active power (kW) | Lower limit of active power (kW) | Upper limit of reactive power (kvar) | Lower limit of reactive power (kvar) | Capcity(kW) |
|---|---|---|---|---|---|---|
| VSC1 | $U_{dc}Q$ | 2000 | -2000 | 2000 | -2000 | 2000 |
| VSC2 | PQ | 2500 | -2500 | 2500 | -2500 | 2500 |

### 4.2 Simulation results and analysis

The configuration results of the two-stage optimal model considering the operation characteristics of AC/DC hybrid distribution system are shown in Table 3. Three typical scenarios are set in the case study, namely, the typical daily scenario in winter and the summer, and the heavy loading scenario during Winter Olympic game. It worth noting that the first two scenarios are used for the optimal configuration of SESS in the first stage, and the third scenario is used for the optimal configuration of collaborative operation of MESS and SESS. A set of comprehensive comparative studies have been carried out in this section to analysis the techno-economic impacts of the BESSs under different scenarios considering the coordinated operation of the SESSs and the MESSs, as well as the thermal management.

*Table 3. Configuration results of the two-stage BESS optimization model*

| ESS | $P_{rate}$ (kW) | $Q_{rate}$ (kvar) | $E_{rate}$ (kW·h) | SOC0(%) | Node |
|---|---|---|---|---|---|
| SESS1 | 500 | 500 | 3500 | 10 | 20 |
| SESS2 | 1100 | 1100 | 6000 | 10 | 13 |
| MESS1 | 3000 | 3000 | 3000 | 30 | 20 |
| MESS2 | 3000 | 3000 | 3000 | 50 | 13 |
| MESS3 | 1000 | 0 | 1000 | 50 | 4 |

**First stage: SESS optimization**

At the first stage, the SESS is optimized based on two typical days in winter and summer, and their LCC is shown in Table 4. where the total cost of the SESS configuration is 326,894.8 $/year and the initial investment is 1,709,200$. The arbitrage income and loss reduction income in winter are 183,861.5$/year and 11,146.9$/year respectively. Since the heavy loading conditions in this area are mostly at peak time, and the load profile in summer is less than winter, the voltage problem is prominent, the arbitrage income and loss reduction are less than those in winter, which are 157,123.1$/year and 1,620.5$/year respectively. Combined with the LCC and various incomes of SESS, the annual net income of SESS is 26,857.2$/year. It can be seen that the income of SESS mainly comes from arbitrage income. The maximum pricing difference between peak and valley in this region is 0.152$/(kW·h), which make it possible for SESS to obtain net income. The peak valley electricity pricing difference can improve the economy of the BESS and promote the development and implementation.

*Table 4. LCC of SESS*

| SESS | Annual capital investment($/year) | Annual fixed operation and maintenance cost($/year) | Annual variable operation and maintenance cost($/year) | Annual replacement cost($/year) | Annual disposal cost($/year) | Total LCC($/year) |
|---|---|---|---|---|---|---|
| SESS1 | 72,915.3 | 11,923.1 | 1,006 | 25,762.2 | 4,433.6 | 116,040.2 |
| SESS2 | 127,850.3 | 26,230.8 | 1,756 | 45,263.6 | 9,753.9 | 210,854.6 |

According to Fig. 6, the load in this area is heavy in the morning and evening at winter season, the voltage profiles of AC system 1 and AC system 2 exceeds the limit before configuring SESS. Moreover, due to the large PV power injection of AC system 1 between 12:00 and 13:00, the voltage of 14 nodes exceed the upper limit. The voltage can be effectively managed within the safety margin through the utilization of SESS. According to Fig. 7(a) and Fig. 8(a), SESS1 is charged at 13:00-14:00 and SESS2 is charged at 12:00 to absorb excessive PV power injection to keep the voltage level within the upper limit, otherwise the two SESSs are in arbitrage operation. In addition, the two SESSs mainly provide reactive power compensation locally in other periods to adjust the voltage apart of the SESS2 reactive compensation during 12:00-13:00.

Due to the low ambient temperature in winter, self-heat release through battery operation is not sufficient to control the temperature, so air conditioning is required to participate the heat management. The operating power of each SESS air conditioner is shown in Fig. 7(b) and Fig. 8(b), and the changes of internal temperature of each SESS and battery surface temperature are shown in Fig. 7(c) and Fig. 8(c) where the battery surface temperature during charging/discharging is slightly higher than the internal temperature of SESS. The operating power of VSC is shown in Appendix Fig. III.

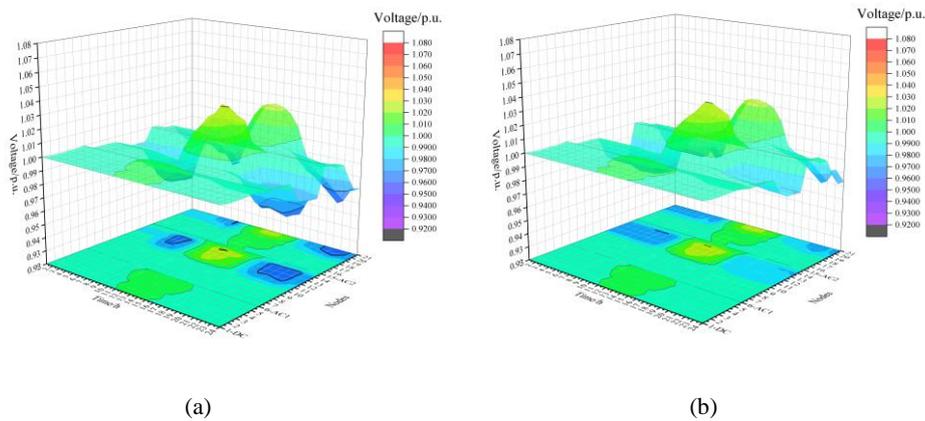

(a)　　　　　　　　　　　　　　　　　(b)

*Fig. 6 Voltage profiles of each typical node (a) before and (b) after SESS configuration in winter*

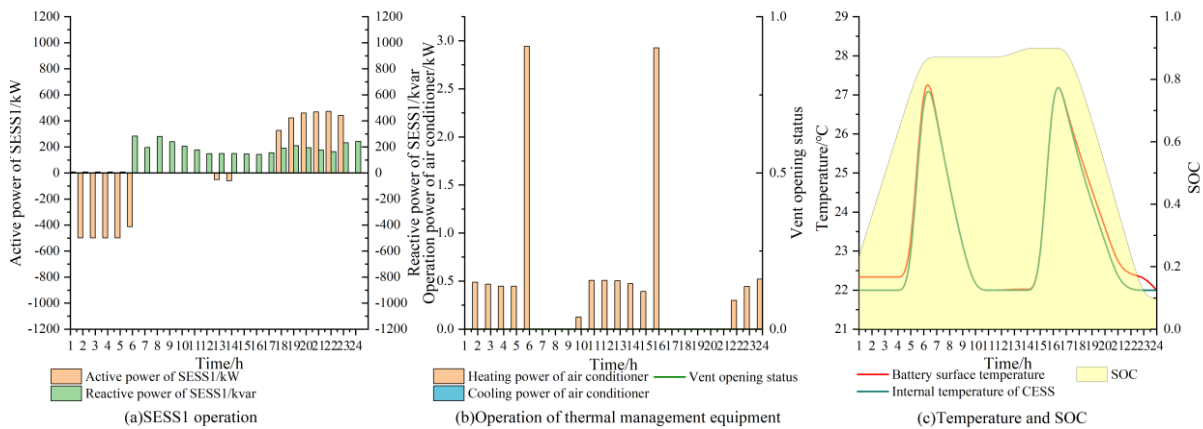

(a)SESS1 operation　　　　　　(b)Operation of thermal management equipment　　　　　　(c)Temperature and SOC

*Fig. 7 Operating state of SESS1 under winter scenario*

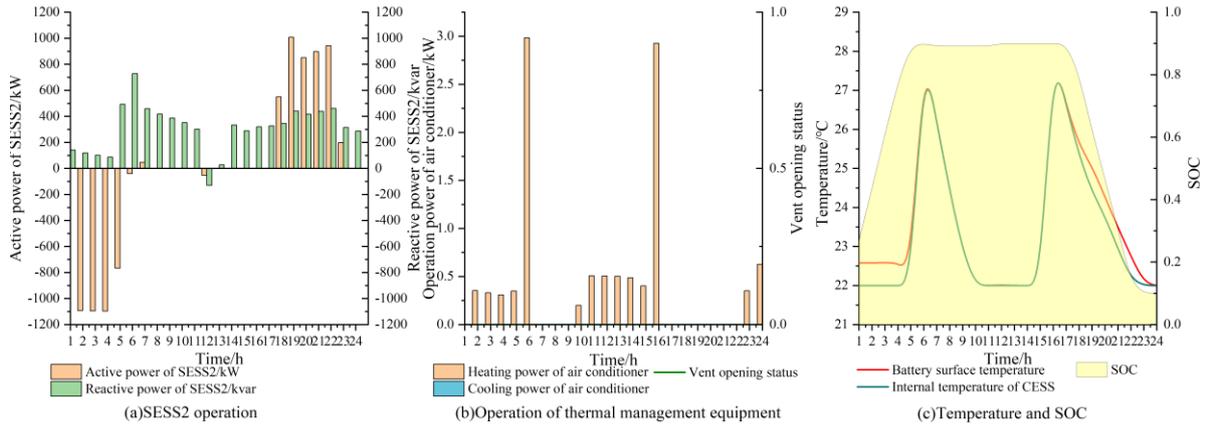

*Fig. 8 Operating state of SESS2 under winter scenario*

The loading condition in this area is light in summer and the PV output is larger than winter, the voltage profiles of the AC system1 exceeds the upper limit during 9:00-16:00 and AC system2 exceeds the upper limit during 10:00-16:00, but the situation can be eased through the utilization of SESS. As shown in Fig. 10(a) and Fig. 10(b), under the period of large PV injection, the SESS not only balance the active power, but absorb reactive power to adjust the voltage within the safety margin. At other times, the SESS is in arbitrage operation and reactive power compensation is carried out locally. Different from the winter scene, the temperature in this area is relatively cool during summer. At most of the time, the internal temperature of the SESS can be managed within the required range through the vent, hence the electricity bill on air conditioning can be reduced. The internal temperature of SESS is maintained by air conditioning at only a few periods of time. The air conditioner operation curve and vent opening state of each SESS are shown in Fig. 10(b) and Fig. 11(b). The changes of internal temperature of each SESS and battery surface temperature are shown in Fig. 10(c) and Fig. 11(c). Comparing the corresponding results shown in Fig. 10(a), Fig. 11(a), Fig. 10(c) and Fig. 11(c), it can be seen that with the increase of battery operating power, the difference between battery surface temperature and internal temperature of SESS increases, and vice versa.

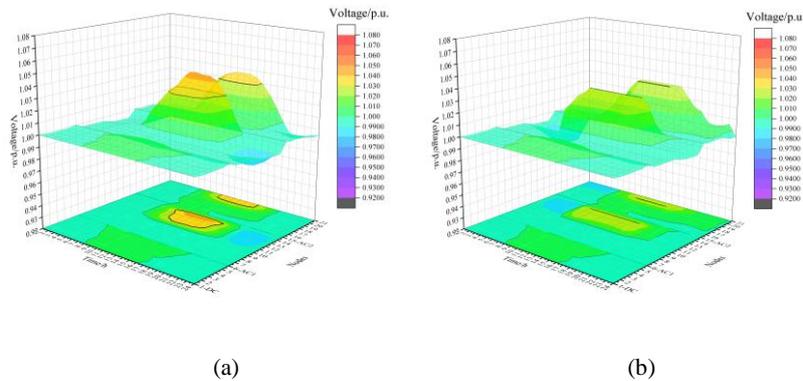

(a)          (b)

*Fig. 9 Voltage profiles of each typical node (a) before and (b) after SESS configuration in summer*

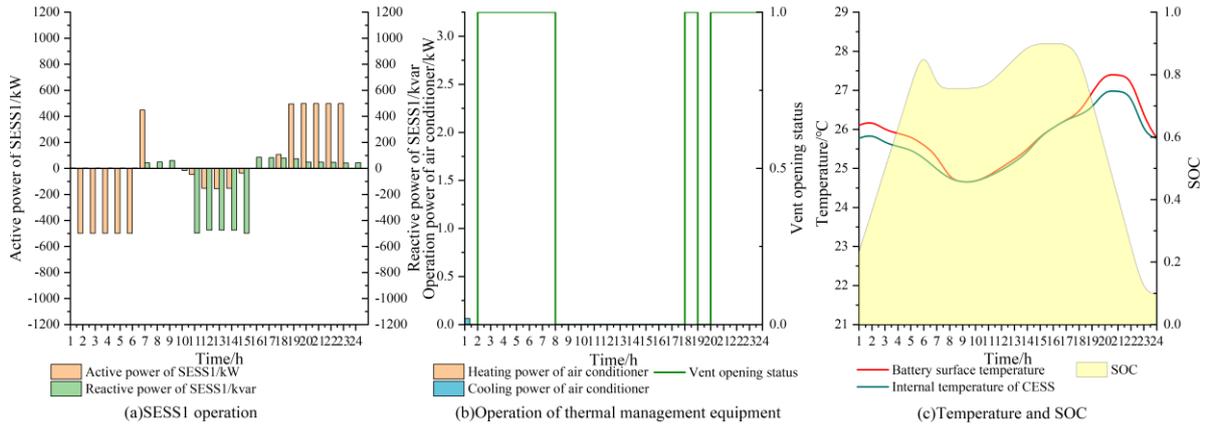

*Fig. 10 Operating state of SESS1 under summer scenario*

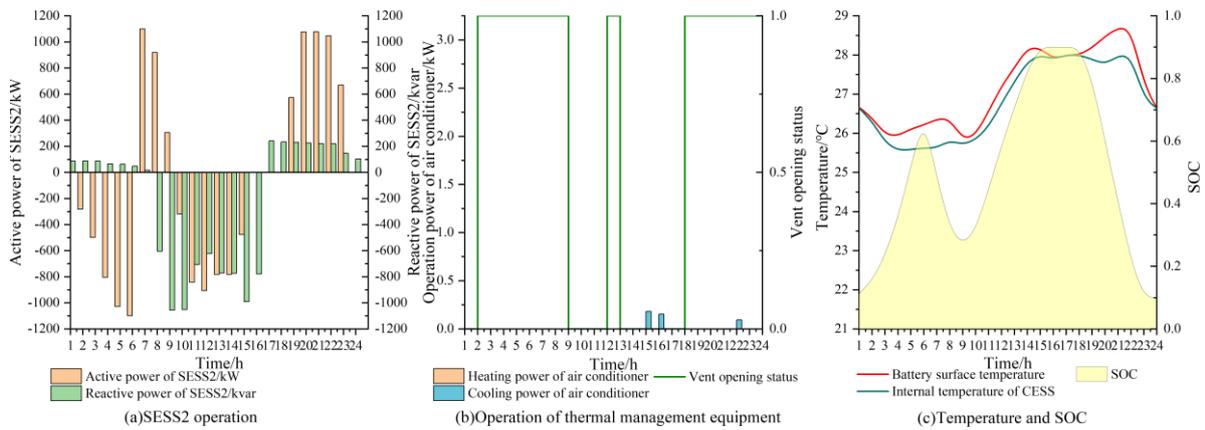

*Fig. 11 Operating state of SESS2 under summer scenario*

**Second stage: MESS optimization**

At the second stage, the MESS under heavy loading condition during Winter Olympics event is organized based on the configuration scheme of SESS at the first stage. The configuration results are shown in Table 3. Under this configuration scheme, the cost-effectiveness of the MESS and the SESS is shown in Table 5. According to Table 5, the rent fees are approximately 68.3%, 68.7% and 67.3% of the overall cost of the MESS which are considered as the majority expenses among other cost. Under the Winter Olympics scenario, the total cost of MESS and SESS is 96,412.4$ and total income is 91,515.5$. The main income of BESS at this stage still comes from arbitrage revenue. And the net cost after deducting income is 4,896.9$.

*Table 5. Cost-benefit analysis of MESS and SESS under Winter Olympic scenario*

| BESS | Rent fee($) | Fixed operation and maintenance cost($) | Variable operation and maintenance cost($) | Arbitrage income($) | Loss reduction income($) | Total cost($) |
|---|---|---|---|---|---|---|
| SESS1 | \ | 1,987.2 | 280 | | | |
| SESS2 | \ | 4,371.8 | 411 | | | |
| MESS1 | 26,140 | 11,923 | 215.3 | 86,027 | 5,488.5 | 4,896.9 |
| MESS2 | 26,567 | 11,923 | 205 | | | |
| MESS3 | 8,342.3 | 3,974.4 | 72.4 | | | |

It can be seen from Fig. 12(a)-(c) that the AC/DC hybrid distribution system suffers serious voltage contingencies at 5:00-9:00 and 15:00-24:00 due to heavy loading conditions without BESS configuration. The situation can be eased to a certain extent with SESS, and can be completely managed with collaboration of SESS and MESS.

According to the BESS operation shown in Figure 13, each BESS has different degrees of discharging under heavy loading conditions at 6:00-8:00 and 16:00-22:00 to improve the voltage level of each node. Due to the excessive PV output during 11:00-13:00, the comprehensive loading level of the AC/DC hybrid distribution system is relatively low. The BESS of each AC system start to charge during this period to adjust the SOC and ensure the normal operation in subsequent periods. Each MESS is charged between 23:00-24:00 to ensure that SOC in subsequent periods meets the requirements of emergency power supply of important loads. Due to the heavy load in this scenario, each BESS generates reactive power in each period to support the voltage locally. The MESS is responsible for the emergency power supply of important loads, so the minimum SOC of the MESS is larger than that of SESS. The Winter Olympic game venues and infrastructure are challenged by the low ambient temperature, the operation state of each CESS air conditioner works at the similar way with the first stage winter scenario. And the higher the operating power of the battery, the higher the surface temperature is comparing to the internal temperature of the CESS.

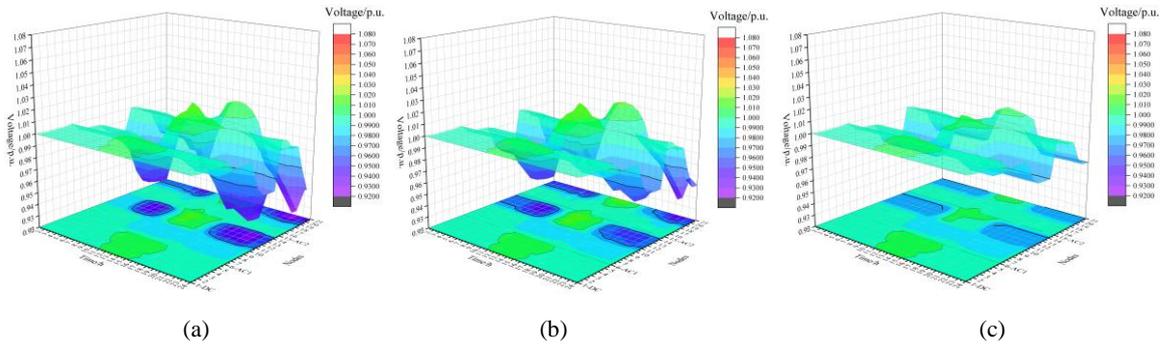

(a)                        (b)                        (c)

*Fig. 12 Voltage profiles of each typical node before and after MESS and SESS configuration under Winter Olympic scenario*
*(a) Voltage profiles without MESS and SESS configuration, (b) Voltage profiles with SESS configuration only, (c) Voltage profiles with MESS and SESS configuration*

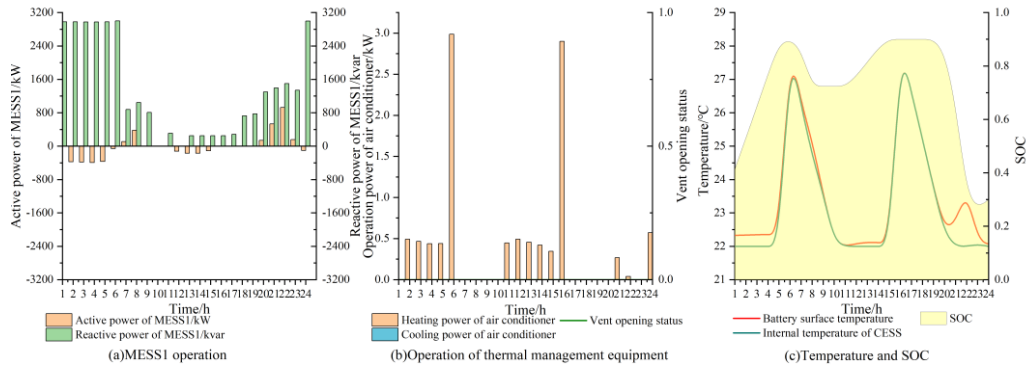

(a) MESS1 operation        (b) Operation of thermal management equipment        (c) Temperature and SOC

*1) Operating state of MESS1*

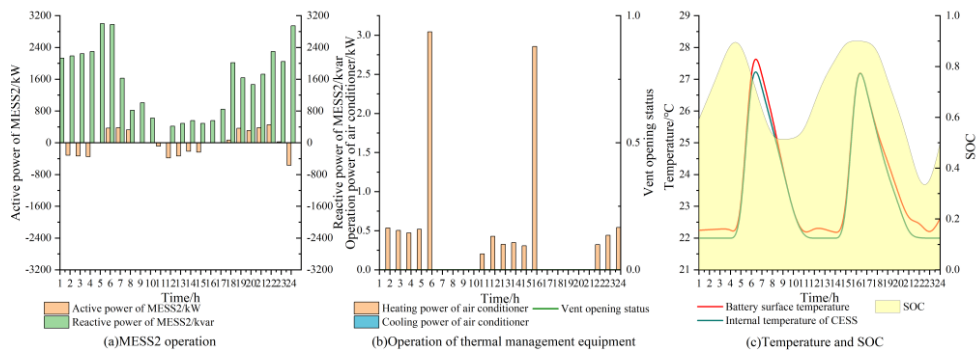

(a) MESS2 operation        (b) Operation of thermal management equipment        (c) Temperature and SOC

*2) Operating state of MESS2*

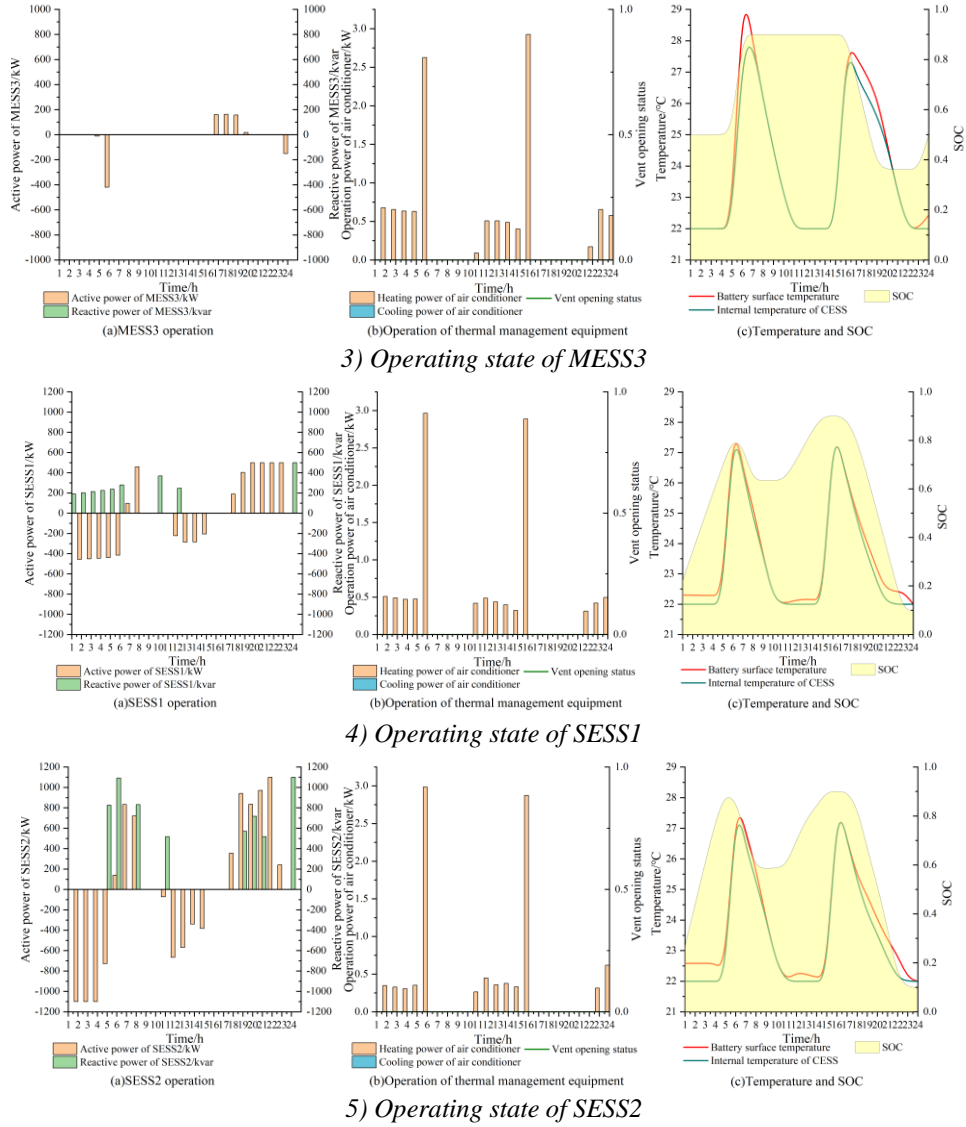

*3) Operating state of MESS3*

*4) Operating state of SESS1*

*5) Operating state of SESS2*

**Fig. 13** Operating state of each BESS under Winter Olympic scenario

It can be seen from Table 5 and Fig. 12 that the configuration of MESS can assist the SESS to solve the problem of short-term voltage contingencies, and the configuration cost of MESS can be reduced through its arbitrage operation. Combined with the operation results of the first stage, the cost of leasing the MESS is higher than that of configuring SESS, and cannot achieve net income. However, the initial investment for SESS is significant. Within the budget, the mixed allocation and operation of SESS and MESS can meet the requirement for mega-event in the short term and improve the economy of the overall system.

## 5. Conclusions

In order to meet the requirements of carbon intensive mega-event while ensuring high operation reliability, a joint SESS and MESS optimal configuration method of AC/DC hybrid distribution system that make full consideration of the thermal management of BESS has been developed in this paper. The following conclusions can be drawn:

1) A CESS thermal management model is established. Through the utilization of air conditioner and natural ventilation onboard of CESS, the temperature of CESS can be managed within the optimization range, and the natural heat exchange through the vent can reduce the electricity bill of air conditioner and operation cost.

2) A refined and comprehensive BESS LCC model is proposed. The variable operation and maintenance cost including self-loss cost of battery and thermal management cost of CESS are included to accurately track the BESS performance from a life cycle angle.

3) A collaborative operation scheme of active/reactive power of SESSs and VSCs in AC/DC hybrid system is presented. The cooperation scheme can improve the system acceptability of REGs by smoothing the fluctuations and avoiding curtailment without additional infrastructure upgrade. At the same time, the linearized battery life cycle model and thermal management model can guide the economic operation of the system.

4) A coordinated operation strategy of SESS and MESS is proposed to satisfy the requirements of various scenarios. The SESS can gain revenue through arbitrage and auxiliary services to compensate the high capital investment, the shared MESS can be an essential complement to ensure the reliability during the mega-event without additional infrastructure investment.

5) A two-stage optimization model that manages the charging/discharging activities and configuration capacity of the SESSs from a life cycle point of view and the MESSs from the perspective of scenario cycle cost angle is developed. The collaborative operation of SESS and MESS can solve the main issues during the mega-event under normal and abnormal conditions. The SESS can effectively maximize the revenues through arbitrage and auxiliary services while preventing severe degradation of batteries after the event.

*Acknowledgements*

This work is partially supported by the National Natural Science Foundation of China (51777134, 52061635103) and the National Key Research & Development Program (2016YFB0900500 and 2016YFB0900503).

*Appendix*

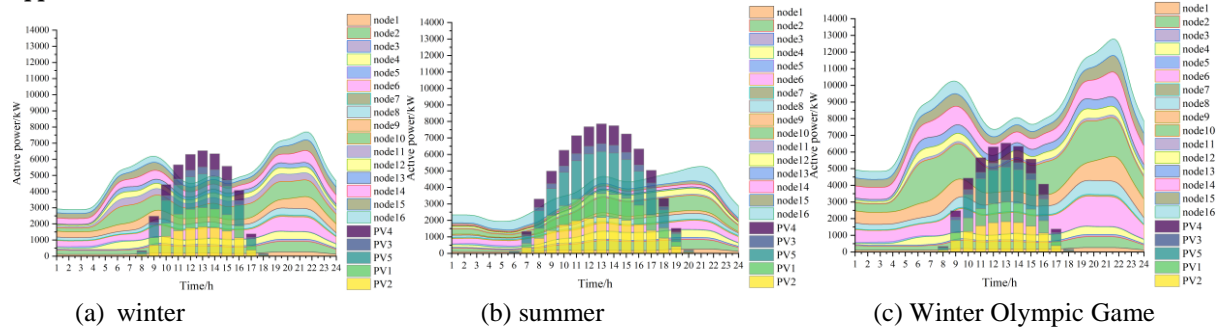

(a) winter      (b) summer      (c) Winter Olympic Game

*Fig. Ⅰ Daily load and PV profiles under different scenarios*

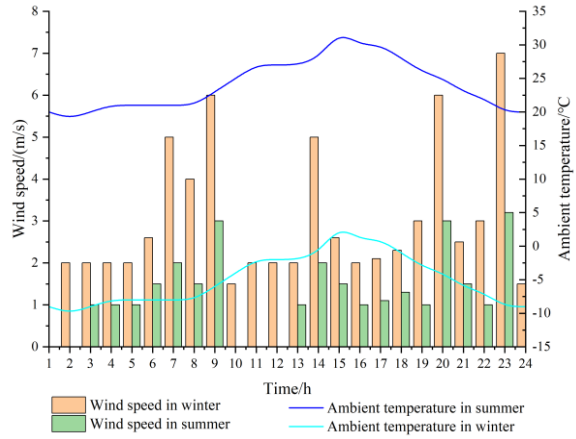

*Fig. II* The wind speed and ambient temperature for each typical scenario

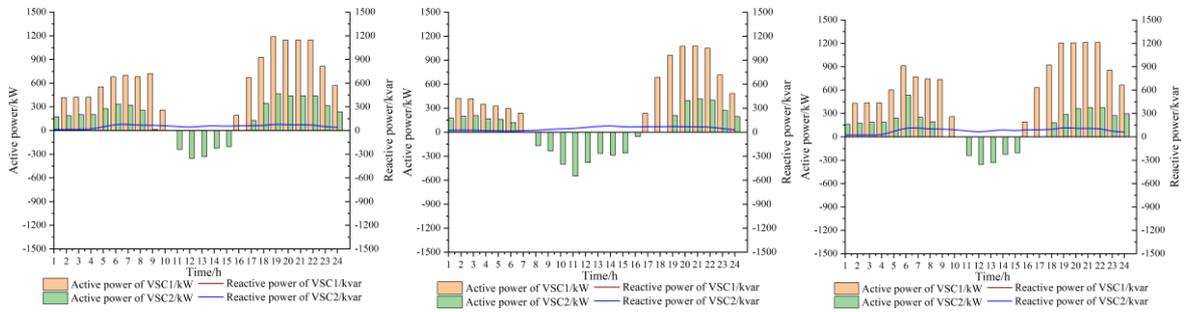

(a)VSC operation power in winter (b)VSC operation power in summer (c)VSC operation power in Winter Olympic scenario

*Fig. III* The VSC operation power under typical scenarios

*Nomenclature*

| Abbreviation | | $\eta_{loss}^{VSC}$ | power losses coefficient of VSC |
|---|---|---|---|
| BESS | battery energy storage system | $\eta_{loss}^{VSC}$ | active loss coefficient of the VSC |
| BMS | battery management system | $\delta$ | self-discharging rate of battery |
| CESS | container energy storage system | $\Omega_i$ | set of adjacent nodes of node $i$. |
| MESS | mobile energy storage system | $\tau$ | discount rate (%) |
| PCS | power conversion system | $\rho^{air}$ | air density |
| REG | renewable energy generations | $\lambda_1 / \lambda_2$ | weight coefficients |
| SOP | the soft open points | $\phi_i$ | set of receiving end nodes that receive power from node i |
| SOC | state-of-charge | $\varphi_i$ | the set of sending end nodes that have receiving end nodes i |
| SESS | stationary energy storage system | $\mu_{impor,i}$ | important load ratio |
| TOU | time-of-use | **Variables** | |
| **Parameters** | | $B_{arb}$ | arbitrage revenue of BESS |
| $A_{bar}$ | heat dissipation area of a single battery | $C_{pun}$ | cost of punishment applied by utility for unsatisfied relaxation deviation issues |
| $A_{vent}$ | vent area | $C_{com,\omega}$ | compensation cost of battery life expenditure |
| $A_{wall}$ | area of CESS wall | $C_{loss0} / C_{loss}$ | network loss cost before/after installing BESS |
| $c_E$ | price of SESS capacity per kW·h | $DOD_{ref,\omega}(i)$ | depth of discharge of i-th charging and discharging cycle in scenario ω |
| $c_P$ | PCS cost per kW | $DOD_{lin,j}(t)$ | a cycle depth of discharge |
| $c_B$ | cost of the necessary supporting facilities per kW·h | $E_{rate,i}$ | rated capacity of i-th BESS |
| $c_f$ | fixed operation and maintenance cost per kW | $I_{ij,ac,\omega}(t)$ | current flowing from AC node $i$ to $j$ |
| $c_d$ | SESS specific disposal cost per kW of the battery | $I_{ij,dc,\omega}(t)$ | current flowing from DC node $i$ to $j$ |
| $c_{rent}$ | daily rent of MESS unit modul | $I_{bat,i,\omega}(t)$ | current of single battery for CESS |
| $C_{Spe}^{bat}$ | specific heat of battery | $k$ | replacement times of battery |
| $C_{Spe}^{air}$ | specific heat of air | $n$ | average lifetime of the chosen SESS |
| COP/EER | energy efficiency ratios of air conditioning heating/ refrigeration | $N_{CESS,i}$ | number of CESS constituting BESS |
| $C_{flo}$ | discharge coefficient of the openning | $N_{MESS,i}$ | number of MESS unit module |
| $C_{wind}$ | wind effect coefficient | $N_{DOD}$ | number of charging and discharging cycles in a day |
| $C_{bud}$ | initial investment budget of the project | $P_{ch,i,\omega}^{BESS}(t) / P_{dis,i,\omega}^{BESS}(t)$ | charging / discharging active power of i-th BESS at time t in scenario ω |
| $D_\omega$ | number of days under scenario ω | $P_{rate,i}^{BESS} / Q_{rate,i}^{BESS}$ | maximum active power/maximum reactive power of i-th BESS |
| $E_{rate}^{MESS}$ | rate energy capacity of MESS unit module | $P_{ac,h,\omega}^{VSC}(t)$ | injected power into AC system via h-th converter at time t in scenario ω |
| $E_{SESS}^{min} / E_{SESS}^{max}$ | minimum/maximum investment capacity of SESS | $P_{rate,i}^{SESS}$ | rate power of SESS |
| $gap_{max}$ | max relaxation deviation | $P_{hot,i,\omega}^{Air}(t) / P_{cool,i,\omega}^{Air}(t)$ | heating power/cooling power of each CESS air conditioner of i-th BESS at time t in scenario ω |
| $H$ | number of VSCs | $P_{ij,ac,\omega}(t) / Q_{ij,ac,\omega}(t)$ | active power/reactive power from AC node i to AC node j in time t |
| $h_{trans}$ | heat transfer coefficient | $P_{i,ac,\omega}(t) / Q_{i,ac,\omega}(t)$ | sum of active power/reactive power injected into AC node i in time t |
| $I_{ij,ac}^{max}$ | maximum current of AC branch ij | $P_{dis,i,ac,\omega}^{SESS}(t) / P_{ch,i,ac,\omega}^{SESS}(t)$ | active power/reactive power of SESS injected into AC node i during t period |
| $I_{ij,dc}^{max}$ | maximum current of DC branch ij | $P_{i,ac,\omega}^{VSC}(t) / Q_{i,ac,\omega}^{VSC}(t)$ | active power and reactive power injected into AC side of VSC |
| $K_{wall}$ | heat capacity | $P_{ij,dc,\omega}(t)$ | active power from DC node i to DC node j in time t |
| $M^{air}$ | quality of air | $P_{i,dc,\omega}(t)$ | sum of active power injected into DC node i in time t |
| $M^{bat}$ | battery quality | $P_{dis,i,dc,\omega}^{SESS}(t) / P_{ch,i,dc,\omega}^{SESS}(t)$ | active power of SESS injected into DC node i during t period |
| $N_{BESS}$ | number of installed BESS | $P_{i,ac,\omega}^{VSC}(t)$ | active power flowing through DC side of VSC |

| Symbol | Description | Symbol | Description |
|---|---|---|---|
| $N_{ac}$ | number of nodes in AC subsystem | $P_{loss,h,\omega}^{VSC}(t)$ | active loss of the h-th VSC |
| $N_{dc}$ | number of nodes in DC subsystem | $P_{rate,i}^{SESS}$ | rated active power of SESS |
| $N_{bar}$ | number of single cells constituting CESS of each container | $Q_{i,\omega}^{BESS}(t)$ | charging / discharging active power of i-th BESS at time t in scenario ω |
| $N_{par}$ | number of parallel branches of battery in each CESS | $Q_{ch,i,\omega}^{BESS}(t)/Q_{dis,i,\omega}^{BESS}(t)$ | charging/discharging reactive power of i-th BESS at time t in scenario ω |
| $N_{MESS,max}$ | maximum number of MESS unit module | $Q_{CESS,i,\omega}^{heat}(t)$ | heat(kW) generated by the normal operation of CESS |
| $price(t)$ | electricity price at time t | $Q_{rel,i,\omega}^{heat}(t)$ | heat released into the air |
| $P_{max}^{Air}(t)$ | maximum operating power of air conditioner | $Q_{abs,i,\omega}^{heat}(t)$ | self absorption heat of the battery |
| $P_{i,ac}^{PV}(t)$ | active power injected by PV on AC node i in time t | $Q_{vent,i,\omega}^{heat}(t)$ | vent heat exchange |
| $P_{i,ac,\omega}^{LOAD}(t)/Q_{i,ac,\omega}^{LOAD}(t)$ | active power and reactive power consumed by load during t period | $Q_{wall,i,\omega}^{heat}(t)$ | wall heat exchange |
| $P_{i,dc,\omega}^{PV}(t)$ | active power injected by PV on DC node i in time t | $Q_{abstem,i,\omega}^{heat}(t)$ | second stage heat exchange of battery temperature change |
| $P_{i,ac,\omega}^{LOAD}(t)$ | active power consumed by load during t period | $Q_{i,ac,\omega}^{SESS}(t)$ | active power and reactive power of SESS injected into AC node i during t period |
| $P_{ac,max,h}^{VSC}/Q_{ac,max,h}^{VSC}$ | upper limit of active power and reactive power transmitted by VSC | $S_i^{BESS}$ | PCS capacity |
| $P_{SESS}^{min}/P_{SESS}^{max}$ | minimum/maximum investment power of SESS | $SOC_{i,\omega}(t)$ | SOC of i-th BESS at time t in scenario ω |
| $R_{ij,ac}/R_{ij,dc}$ | resistance of the AC amd DC branch ij | $SOC_{max,i}/SOC_{min,i}$ | upper/lower limits of SOC of i-th BESS |
| $R_{int}^{bat}$ | internal resistance of single battery for CESS | $SOC_{avg,j}$ | average daily state of charge |
| $S_h^{VSC}$ | access capacity of the h-th VSC | $SOC_{avg,\omega}$ | the average state of charge in scenario ω |
| $T_{rent}$ | rent cycle | $T_{bar,i,w}(t)$ | battery surface temperature |
| $T_{ext,i,w}(t)$ | the external ambient temperature | $T_{CESS,i,w}(t)$ | CESS internal temperature |
| $V_{wind}$ | wind speed | $T_{CESS,min}/T_{CESS,max}$ | Lower/upper limits of CESS internal temperature |
| $V_{i,ac}^{max}/V_{i,ac}^{min}$ | upper/lower limits of the voltage of the AC node i | $T_{i,\omega,avg}$ | average temperature of BESS in the i-th charging and discharging cycle |
| $V_{i,dc}^{max}/V_{i,dc}^{min}$ | upper/lower limits of the voltage of DC node i | $V_{i,dc,\omega}(t)$ | voltage of DC node i at time t |
| $W$ | collection of scenarios | $V_{i,ac,\omega}(t)$ | voltage of AC node i at time t |
| $X_{ij,ac}$ | the reactance of the AC branch ij | $X_{vent,i,w}(t)$ | 0-1 variable indicating the opening status of the vent |
| $Y$ | project cycle (year) | $\mu_{dis,i,\omega}(t)/\mu_{ch,i,\omega}(t)$ | discharging/charging sign of i-th BESS at time t in scenario ω |
| $\alpha$ | average annual decline rate of the SESS capital investment | $\varepsilon$ | times of replacement |
| $\alpha_{vent}$ | volumic flow rate | $\zeta_{lin,j}$ | daily life damage of j-th MESS module |
| $\eta_C/\eta_D$ | charging/discharging efficiency of battery | $\zeta_{lin,j}^{idl}$ | an amount of capacity lost from idling in one day |
| $\eta_{PCS}$ | PCS efficiency | $\zeta_{lin,j}^{cyc}$ | an amount of capacity lost in one cycle |